\documentclass[]{spie}

\usepackage{amsmath, amssymb, amsfonts, mathrsfs, amscd, color, graphicx}
\usepackage{wrapfig}

\newcommand{\R}{\mathbb{R}}
\newcommand{\G}{\mathbb{G}}
\newcommand{\C}{\mathcal{C}}
\newcommand{\sigmoid}{\mathscr{S}}

\newcommand{\g}{g}

\newcommand{\X}{\mathcal{X}}

\newcommand{\cell}{\xi}
\newcommand{\cortex}{\textnormal{cortex}}
\newcommand{\retina}{\textnormal{retina}}

\newcommand{\CMF}{\textnormal{{\small{CMF}}}}
\newcommand{\PI}{\textnormal{{\small{PI}}}}

\title{Geometry and dimensionality reduction of\\ feature spaces in primary visual cortex}

\author{Davide Barbieri\skiplinehalf
Universidad Autonoma de Madrid, 28049 Madrid, Spain
}

\authorinfo{e-mail: davide.barbieri@uam.es. Proceedings of SPIE 9597 \emph{Wavelets and Sparsity XVI} (2015).\\ The author was supported by a Marie Curie Intra European Fellowship (626055, FP7).}

\begin{document}

\maketitle

\begin{abstract}
Some geometric properties of the wavelet analysis performed by visual neurons are discussed and compared with experimental data. In particular, several relationships between the cortical morphologies and the parametric dependencies of extracted features are formalized and considered from a harmonic analysis point of view.
\end{abstract}


\section{Introduction}
The functional behavior of neurons in primary visual cortex can be partially modeled as a linear response to visual stimuli. This linear response can be formally described as a generalized wavelet analysis of images, intended as functions from $\R^2$ to $\R^+$ (we consider greyscale images), with respect to a family of functions $\{\psi_\cell\}_{\cell \in \X} \subset L^2(\R^2)$. Such a family can be indexed by a parameter set $\X$ that encodes some relevant geometric properties, which correspond to local features of the input image, and that is mapped on the cortex. Each neuron can be identified with a given collection of such properties, i.e. by an element of $\X$, and it is said to be \emph{selective} with respect to the associated features. A neuron receives a visual input only from a bounded region in the visual plane. Inside this region, which is called the \emph{receptive field}, a neuron can be selective with respect to several \emph{local features}, such as the local orientation, the local spatial frequency (the rate of oscillations), the local apparent velocity (the component of the speed of motion that is orthogonal to the locally detected orientation). The neurophysiological and computational meaning of this selectivity will be discussed in sections \ref{sec:neuro} and \ref{sec:mathmod}.

The primary visual cortex (V1) of many mammals, notably including humans\footnote{Most of the experimental data reported here involve primates such as humans and macaques, close relatives such as tree shrews, but also cats, and similar behaviors have been encountered in carnivors like ferrets. Notable examples of morphologies that are different from the ones discussed here are provided by rodents\cite{Hooser}.}, shows a remarkably regular spatial organization of the features its neurons are selective to\cite{Hubel}. V1 is arranged in such a way that each feature varies smoothly on the cortical layer: neurons that are close to one another are selective to similar features. Receptive fields are arranged in a topographic, or \emph{retinotopic} way, i.e. neurons that are close to one another in the cortex are associated to similar regions in the visual plane. They are also highly overlapping, in the sense that two adjacent neurons respond to two almost entirely overlapping regions. Orientation varies smoothly, in the sense that two adjacent neurons are selective for similar local orientations, and the same is true for the other listed features. Moreover, the variability of some of the features is mainly associated to the horizontal position of the neurons, according to a so-called hypercolumnar structure whose discovery dates back to the Nobel prizes Hubel and Wiesel. The cerebral cortex can be thought as a thick layered stack of (folded) planes and, roughly speaking, the hypercolumnar paradigm postulates that the function (feature selectivity) of neurons does not vary along the transversal direction but depends only on the position on the cortical plane (see also\cite{Zucker}).

\begin{figure}[ht!]
\centering
\includegraphics[height=.36\textwidth]{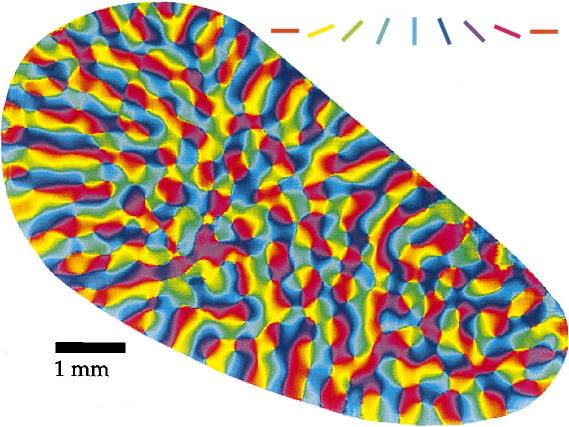} \hspace{80pt}
\includegraphics[height=.36\textwidth]{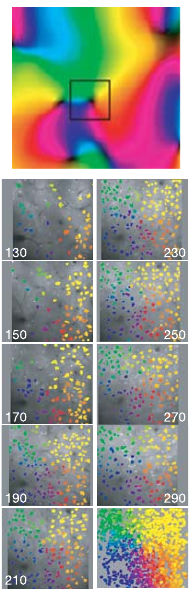}
\caption{Left: a planar representation of a tree shrew V1\cite{Bosking}; colors represent preferred orientations of the underlying neurons. The map is almost everywhere smooth with almost-periodic displacement of topological singularities, around which all orientations are represented. Right: single neuron measurements of preferred orientation at various depths in cats\cite{Ohki}. The feature of local orientation depends only on the horizontal position and does not change transversally.}\label{fig:ohki}
\end{figure}

This structure has led to the representation of the \emph{functional architecture} of primary visual cortex in terms of maps from the cortical plane to the space of a given feature, called \emph{cortical maps}, which describe the displacement of the neurons in terms of their selectivity\cite{Rosa}. Probably the best known cortical map is that of orientation preferences, depicted in Figure \ref{fig:ohki}. It is a map $\Theta : \R^2 \to [0,\pi)$ that indicates, for each point on the cortex, what is the orientation the underlying neuron (population) is selective to. This means that if a light stimulus is present in the receptive field of a neuron, and that stimulus is mainly oriented along the depicted direction (e.g. it contains an edge with that orientation), then the neuron will be maximally stimulated. Conversely, if the stimulus contains an edge in the orthogonal direction, the neuron is minimally stimulated.

A satisfactory understanding of the properties of the family $\{\psi_\cell\}_{\cell \in \X}$ or of the cortical representation of the set $\X$ from the point of view of image processing is still an open problem, even if many partial results are available. This issue is relevant for artificial vision, because the capabilities of analyzing, compressing, and processing images possessed by the visual cortex still outperform any artificial system. It is also relevant for neuropsychology and cognitive studies, because it is still not clear what is the role of the different constituents of the brain in order to build the structured and organized visual representation generated by the visual cortex. On the other hand, this approach suffers from several limitations: just to mention a few, it does not include the various nonlinear mechanisms that contribute to the neural responses, and it does not take into account the dynamics of the network of connected neurons as well as the dynamic of activation of each single neuron.

The purpose of this paper is to review some key facts about mathematical models of neurophysiological behaviors, to show some new properties of these models, to present some open problems that can be of some interest to the community working on harmonic analysis and approximation theory, and to provide some relevant terminology and references commonly used in the neuroscience literature.

\section{A very brief summary of some V1 cellular behaviors}\label{sec:neuro}
In this section we will introduce the main features of V1 that we will consider. We will focus on linear behaviors and on cortical maps, disregarding the several issues on nonlinear behaviors, such as normalizations or surround effects\cite{FitzpatrickBeyond, Carandini2005, CarandiniHeeger}, on dynamical effects, such as spatio-temporal behaviors of receptive profiles\cite{DeAngelis1995, CBS}, neural oscillations\cite{Singer}, neurodynamics\cite{Bojak, Coombes, Kilpatrick, Bressloff, HRH}, or on properties of connectivities\cite{Bosking, Angelucci, AngelucciBressloff} and their geometric models \cite{CS, BCCS}.

\emph{Primary visual cortex} V1 is the first and largest cortical area dedicated to the processing of visual stimuli. It is located in Brodmann area 17, involving both hemispheres of the occipital lobe of the brain, and it is also called striate cortex. It is followed by higher visual areas, from V2 to the so-called V5/MT, also called \emph{extrastriate cortices}, that are located in Brodmann areas 18 and 19. V1 receives direct sensory input from the \emph{lateral geniculate nucleus} LGN, which is located in the thalamus and acts as a preprocessing unit of the visual stimuli collected by the retina. The $\sim 10^8$ retinal receptors are connected by $\sim 10^6$ optic nerve fibers to the LGN and finally to the $\sim 5 \times 10^8$ V1 cells\cite{Valois, Teller}. On the one hand, this gives a strong indication that alredy in the retina there must occur a cospicuous processing of information. On the other hand, the $10^6$ units of infomations reaching V1 are evidently highly reprocessed. V1 transmits its information to extrastriate cortical areas by connections to V2, which is then connected to higher visual areas. The electric inputs that V1 receives are due to three main classes of \emph{connectivities}: \emph{feedforward} inputs coming from the eye, through the lateral geniculate nucleus; \emph{lateral} inputs coming from other V1 neurons; and \emph{feedback} inputs coming from higher cortical areas. V1 neurons are characterized by their feature selectivity, by their physical displacement, and by their connectivities.

\subsection{Classical receptive fields}
Just like all neurons, the neurons that populate V1 accumulate an electric potential on their boundary membrane by means of chemical mechanisms involving ion concentrations. When this electric potential exceeds a given threshold, the neuron releases an impulsive discharge called \emph{action potential}, that is transmitted by synapses as an electric current. Such current flows through connections until it reaches other synapses, where it is again converted into an electric potential that accumulates in the neuron that receives the signal. A neuron that emits an action potential is said to be \emph{firing}, the action potential itself being called a spike. The activity of a neuron is its temporal sequence of spikes, and it is generally quantified by a nonnegative scalar quantity that measures its average \emph{firing rate}, that is the frequency of spikes per second that it emits\cite{Dayan}. A neuron has always a positive firing rate: even when there is no external stimulus acting on the network, each neuron fires from time to time (rest state). The \emph{activity} of a neuron is then measured as the difference between its firing rate and the firing rate at rest\footnote{This does not account for all the information that is handled by the neural systems, and other quantities may be involved such as subthreshold activities\cite{Gerstner}.}. The firing rate depends in a nonlinear way on the membrane potential, which is generally modeled as a sigmoid function; in particular, each neuron has a maximum firing rate, a linear range of membrane potentials where the firing rate is linear, and a minimum firing rate equal to zero.

A V1 neuron's visual receptive field is defined as the region of the visual field where the presence of light (i.e. a visual stimulus) causes a neural response. The direct\footnote{We are disregarding here the activity due to lateral and feedback connections.} activity response of a neuron to a stimulus does not depend only on the stimulus location on the visual field, but also on the spatio-temporal distribution of the light in the receptive field region. One of the basic needs of a neurophysiologically based formal theory of visual perception is to have a quantitative description of such dependence, that is sufficiently accurate to allow to predict the response of a neuron to any chosen visual stimulus. The simplest (though very effective) model is the \emph{Linear-Nonlinear-Poisson} model. Even if it does not account for many nonlinear and/or collective phenomena\cite{Carandini2005}, it is able to reproduce much of the single-neuron behavior. It consists of three stages: 
\begin{enumerate}
\item The visual stimulus produces a feedforward input that feeds a neuron $\cell$ with a voltage: this voltage is considered as depending linearly on the visual stimulus. If $f : \R^2 \to \R^+$ is a visual stimulus, then the resulting membrane potential is a linear map $V : f \mapsto \R$ generally realized as an $L^2(\R^2)$ linear functional
$$
V_\cell[f] = \langle f, \psi_\cell\rangle_{L^2(\R^2)} 
$$
where the vector $\psi_\cell$ depends on the specific neuron.
\item Activity is computed by applying to $V_\cell[f]$ a static nonlinearity, generally a sigmoid
$\sigmoid(r) = \displaystyle\frac{1}{1 - e^{-a(r - b)}}$.
\item The firing process is considered as a Poisson process whose rate is given by $\mathscr{R}_\cell[f] = \sigmoid(V_\cell[f])$. This is equivalent to assume that no information is encoded in the time process of spikes.
\end{enumerate}

The fundamental objects of the LNP model are then the filters $\{\psi_\cell\}_{\cell \in \X}$, where $\X$ represents the family of visual neurons. They are generally called \emph{receptive profiles}, or also \emph{classical} receptive fields\footnote{The expression ``receptive field'' can be found in the literature to indicate either a filter $\psi_\cell$ or its support.}. Other effects, which take place in outer regions and/or that contribute nonlinearly to the receptive profile behavior, are referred to as \emph{nonclassical} or \emph{extra-classical}\cite{FitzpatrickBeyond, Carandini2005}.

In order to describe quantitatively the receptive profiles, one needs to set up an experimental framework that allows to measure them. The technique used is borrowed from a technique introduced by Wiener, that is called \emph{white noise analysis}, or \emph{spike-triggered average} or, more commonly, \emph{reverse correlation}. It basically consists of showing random sequences of stimuli while performin electrophysiological recordings of the spike trains produced by a neuron, and performing averages over the stimuli that preceded a spike. In this way one identifies the type of stimuli that generate the highest linear response of the neuron. This in general is not sufficient to describe the complete behavior, which is more properly described in terms of so-called \emph{Wiener series} including nonlinearities of different types. The procedure is however made consistent by imposing the LNP model to this stochastic analysis\cite{Simoncelli2004}.

\subsection{Cortical maps of feature displacements}

Effective experimental techniques have been developed to visualize large portions of cortical tissue activity in vivo. They are called \emph{optical imaging}, and rely either on \emph{voltage-sensitive dyes} or on \emph{intrinsic signals}\cite{BonhoefferGrinvaldOI}. The first one, developed in the 70's and 80's makes use of dye molecules that are capable to transform changes in the membrane potentials into optical signals. This technique offers a temporal resolution of about $10^{-3}s$ and a spatial resolution of about $10^{-4}m = 100 \mu m$, while the produced signals do not correspond to single cell activity but are due to the averaged electric activity of a local population of neurons (about 250-500 neurons). This technique is considered most often for real-time measurements. The second one, developed in the 90's, is based on the intrinsic changes in the amount of light reflected by brain tissue, which is correlated with the presence of neuronal activity. This mechanism is thought to arise at least in part from local changes in the blood composition that accompany such activity, which can be detected when the cortical surface is illuminated with red light. Active cortical regions absorb more light than less active ones. This technique provides a more accurate spatial resolution of $50 \mu m$ or better, but it provides a very low temporal resolution. However, since the experimental set up is easier, it represents the most frequently used technique to obtain cortical maps. More recently, noninvasive techniques with fMRI have been used to provide evidence of orientation maps in humans \cite{Yacoub}.

\begin{figure}[h!]
\centering
\includegraphics[height=.24\textwidth]{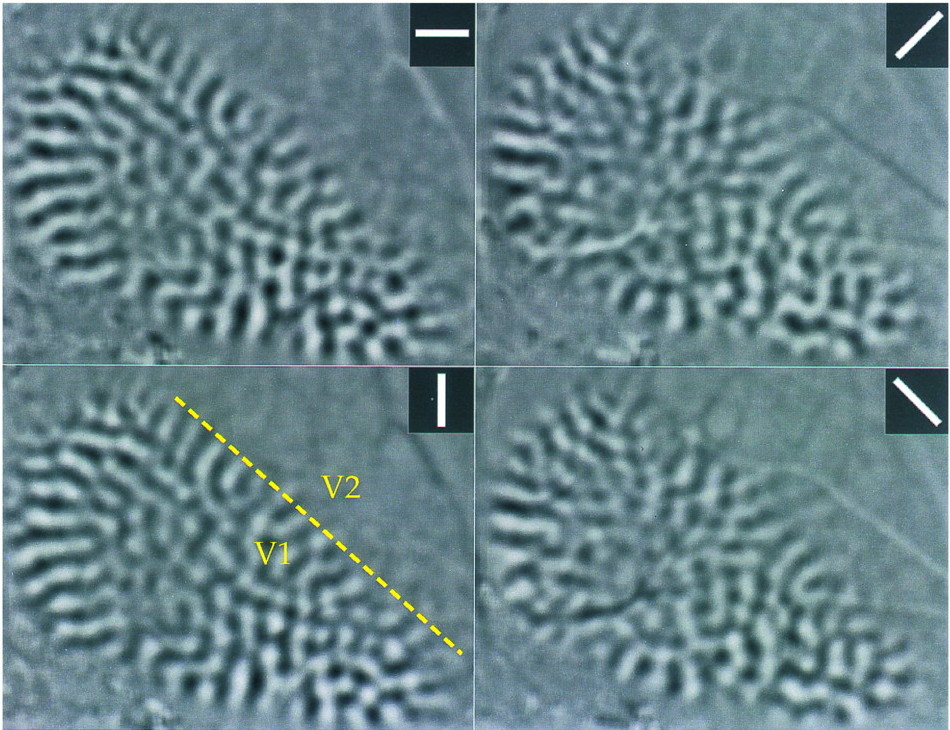} \hspace{30pt}
\includegraphics[height=.24\textwidth]{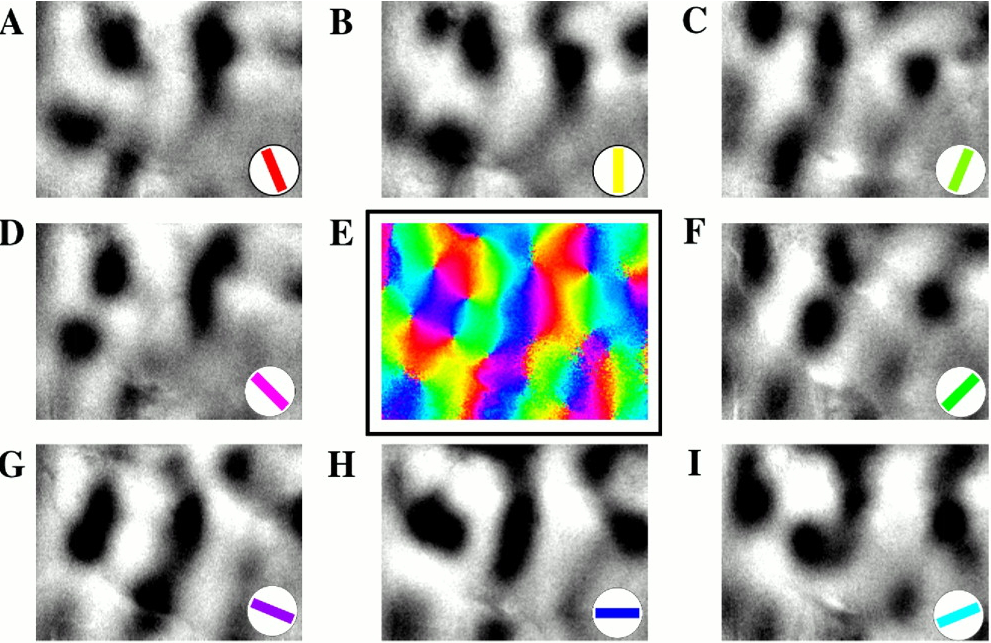} \hspace{30pt}
\includegraphics[height=.24\textwidth]{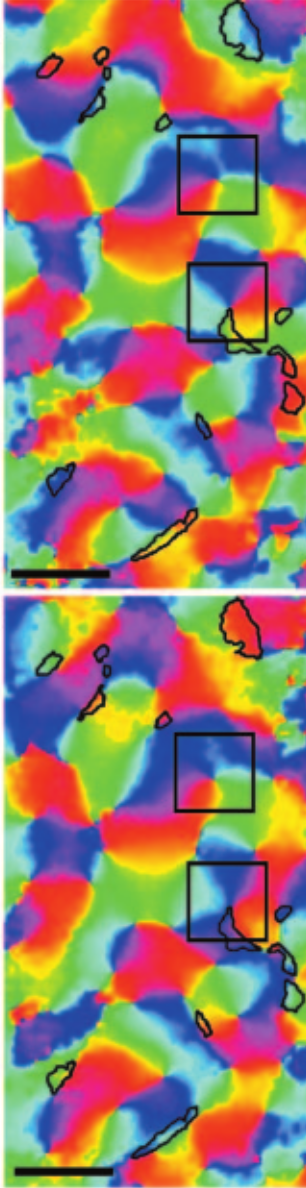}
\caption{Left: orientation-dependent activity maps in tree shrews\cite{Bosking}. Center: vector sum to obtain orientation preference maps  in cats\cite{Crair}. Right: small differences in orientation preference depending on high (top) or low (bottom) spatial frequencies of the presented gratings in cats\cite{Sirovich}.}\label{fig:bosking}
\end{figure}

In Figure \ref{fig:bosking} one can see the experimental data that produce the map of Figure \ref{fig:ohki}. The anesthetized animal is presented with a plane wave (called \emph{grating}) drifting back and forth along a given direction, and the cortical activity is measured in terms of intrinsic signals. One obtains then a collection of maps indexed by a parameter $\theta \in [0,\pi)$, that we can denote with $A:\R^2 \times [0,\pi) \to \R$, where $\zeta \mapsto A(\zeta,\theta)$ represent the local activity around the point $\zeta$ on the cortex with respect to the orientation $\theta$. Observe also that $\theta \mapsto A(\zeta,\theta)$ provides the \emph{orientation tuning} of the neurons located at $\zeta$, i.e. their selectivity curve to different orientations\cite{Webster, Swindale98}. The information about the local orientation preference given by the map $\Theta$ depicted in Figure \ref{fig:ohki} is obtained by the vector sum
$$
\Theta(\zeta) = \frac12 \arg \int_0^\pi e^{i2\theta} A(\zeta,\theta) d\theta .
$$
Since at each $\zeta$ this depends only on one Fourier coefficient of $\theta \mapsto A(\zeta,\theta)$, of course it can not contain all the information given by the function $A$. As a consequence, different activity maps can produce the same orientation preference map. More precisely, since
$$
\Theta(\zeta) = \frac12 \arg \Big(e^{i2\Theta(\zeta)}\int_0^\pi e^{i2\theta} A(\zeta,\theta + \Theta(\zeta)) d\theta\Big) \, ,
$$
then for all $A$ such that the integral at the right hand side is real one obtains the same map $\Theta$. Several experiments have thus focused on the joint displacement of orientation preferences and orientation tuning\cite{Carandini2008}.

Another well studied cortical map concerns the columnar organizations associated to \emph{ocular dominance}\cite{Hubel, Hubener}, that consists of columns where all cells respond preferentially to the same eye. They show a regularly organized structure, which is strictly related to that of orientation preference. However, for the present purposes of the analysis of purely two dimensional visual stimuli, binocularity does not seem to play a central role\cite{Horton}, so we will disregard completely issues regarding ocular dominance.

\newpage

\section{Receptive fields morphologies}\label{sec:mathmod}

This section is devoted to the introduction of some formal properties of classical receptive fields, regarding locations, sizes and shapes, that can be deduced from experimental data.

\subsection{Retinotopy, size and magnification}

\begin{figure}
\centering
\includegraphics[height=.15\textwidth]{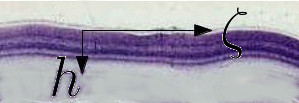} \hspace{20pt}
\includegraphics[height=.15\textwidth]{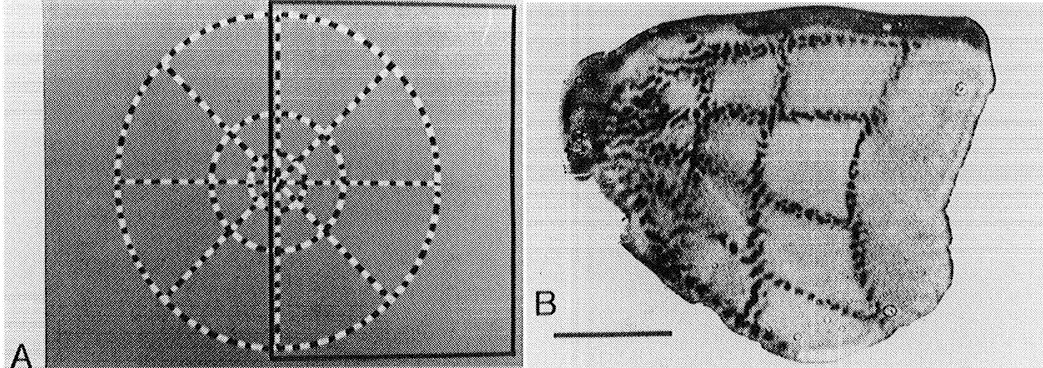}
\caption{Left: coordinates on V1 (edited from \cite{Hubel}). Right: (A) stimulus used for mapping the retinotopic projection onto striate cortex; (B) pattern of response in macaque V1 produced by this stimulus pattern\cite{Valois}.}
\end{figure}

V1 possesses a three dimensional layered structure. We will consider each layer as flat and continuous, and distinguish coordinates on the two dimensional leaves from the transversal coordinate. Some relevant features that will be addressed are indeed approximately constant when measured transversally on V1 \cite{Hubel, Ohki}, while others are commonly considered in average over the layers: for this reason the two dimensional space corresponding to layers is sometimes called the \emph{cortical plane}. Each V1 cell will then be identified by a point
$$
\cell = (\zeta,h) \in V1 = \R^2_{\cortex} \times [0,H] .
$$
We will keep the subscript $\cortex$ in order to distinguish each such plane from the retinal plane $\R^2_{\retina}$, which we will identify with the visual field (this is reasonable only locally: the optical properties of the eye and the retina require spherical coordinates, and for this reason the distances on the visual field are generally measured in angular degrees).
V1 cells can be classified in terms of several properties which concern their functional behavior, typically parameters of their receptive profiles. Let a given property be described by the elements of a set $X$. A \emph{Cortical Map} is then defined as a function
$$m: V1 \to X$$
that associates to a cell at $\cell$ a property $m(\cell) \in X$.

The first properties one is interested to are related to the location and size of receptive fields: if $\Omega_{\cell}\subset \R^2_\retina$ is the receptive field of the cell at $\cell$, denote its center with
$$
q(\cell) = \frac{1}{|\Omega_{\cell}|} \int_{\Omega_{\cell}} x dx \in \R^2_{\retina}
$$
and its size with
$$
S(\cell) = \textnormal{diam}(\Omega_{\cell}) .
$$

Each V1 layer $\R^2_{\cortex}$ is organized in a \emph{retinotopic} way \cite{Hubel, Valois}, i.e. for any fixed $h \in [0,H]$ we have that
$$
q(\cdot,h) = \phi_h : \R^2_{\cortex} \to \R^2_{\retina}
$$
is a smooth bijection. Moreover, for each fixed $\zeta \in \R^2_{\cortex}$ the variation of $q(\zeta,h)$ with $h$, usually referred to as \emph{scatter} \cite{HW1974scatter}, is small and for the present purposes it can be considered negligible, so we will drop the subscript $h$ and simply denote the topographic map as $\phi$. This map is approximately given by the (principal branch of the) complex logarithm, also called \emph{log-polar map} \cite{Valois, GrillSpector}: for $(x,y) \in \R^2_{\retina}$ let $z = x + iy = \rho e^{i\theta} = e^{r + i\theta}$, then
$$
\phi^{-1} : (x,y) \mapsto (r,\theta) = \ln(z).
$$
The map $\phi$ is conformal, i.e. it preserves angles, because $\frac{\partial \phi}{\partial \zeta} \neq 0$, and
its geometry can be understood by noting that rotations and exponential dilations are mapped as translations:
for $\alpha \in S^1$, and $t \in \R^+$
$$
\left\{
\begin{array}{rcl}
\phi^{-1}(e^{i\alpha} z) & = & (r, \theta + \alpha)\\
\phi^{-1}(e^t z) & = & (r + t, \theta) .
\end{array}
\right.
$$
\begin{figure}[t!]
\centering
\includegraphics[height=.13\textwidth]{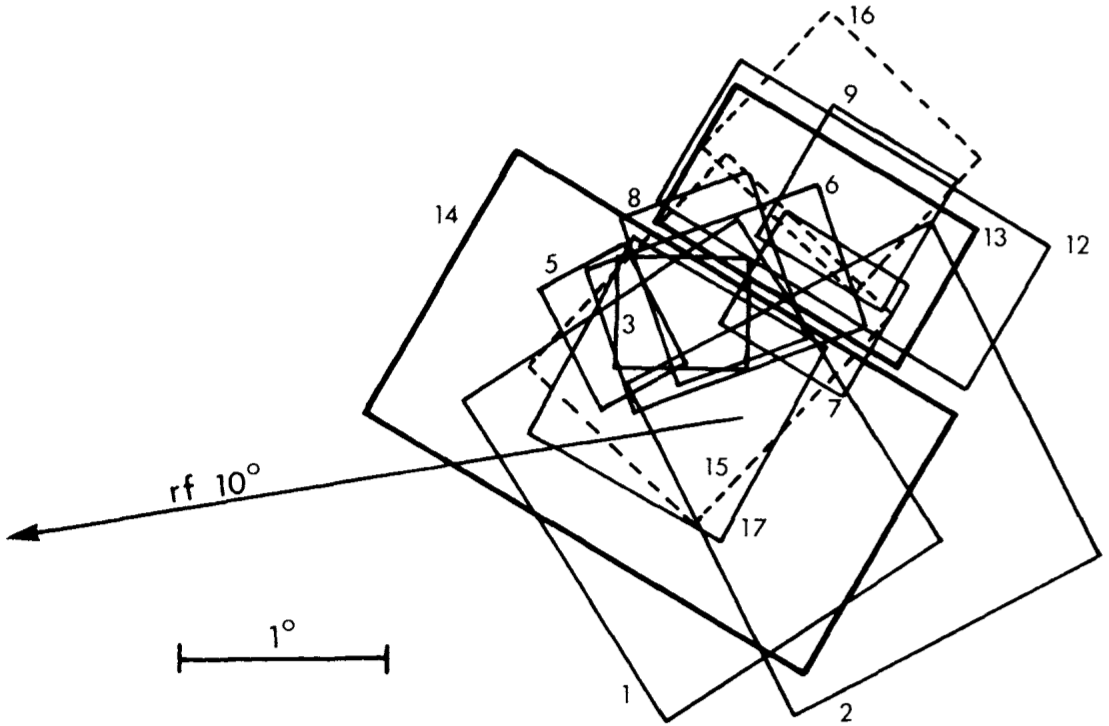}\hspace{70pt}
\includegraphics[height=.13\textwidth]{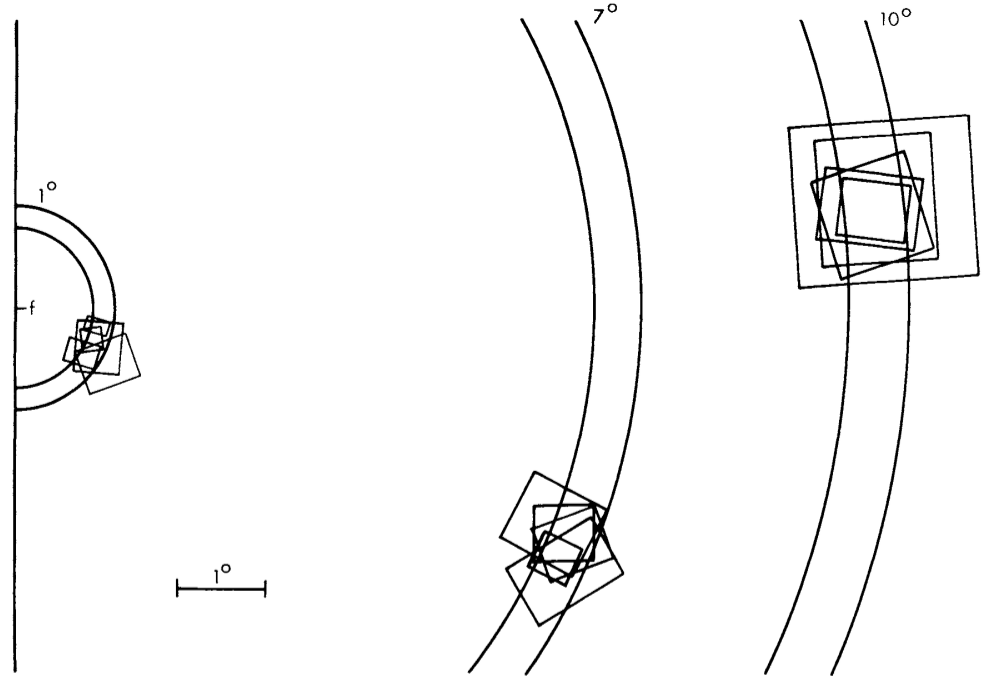}
\caption{Electrophysiological measurements in macaques V1\cite{HW1974scatter}. Left: size and scatter of receptive fields in a transversal penetration, for different depth. Right: receptive field size-plus-scatter and magnification, for three different eccentricities.}\label{fig:hubelwiesel74_scatter}
\end{figure}
A quantity commonly used to quantify properties of the distribution of receptive fields is the \emph{Cortical Magnification Factor} $\CMF$ \cite{DanielWhitteridge1961}. It measures the distance on the cortical surface between two neurons having RF positions whose distance $\delta$ on the visual field is small:\vspace{-6pt}
$$
\CMF_\delta(\zeta_0) = \max_{\zeta \, : \, |\phi(\zeta)-\phi(\zeta_0)| \leq \delta} |\zeta - \zeta_0| .\vspace{-6pt}
$$
Making use of the log-polar mapping and denoting with $z_0 = \phi(\zeta_0)$, since
\begin{displaymath}
\CMF_\delta(\zeta_0) = \max_{z : |z-z_0| \leq \delta}|\phi^{-1}(z)-\phi^{-1}(z_0)| = \max_{z:\,|z - z_0|\leq \delta} \left|\ln\left(z\right) - \ln\left(z_0\right)\right| = \max_{w:\,|w|\leq \delta} \left|\ln\left(1+\frac{w}{z_0}\right)\right| \approx \frac{\delta}{|z_0|}
\end{displaymath}
we get\vspace{-6pt}
\begin{equation}\label{eq:CMF}
\CMF_\delta(\zeta_0) \approx \frac{\delta}{e^{r_0}} = \frac{\delta}{\rho_0}.
\end{equation}
A related notion is that of \emph{point image} $\PI$: it is defined as the square root of the cortical surface area activated by a point in the visual space. It is then approximately proportional to the distance on the cortical surface at which two cells have separate receptive fields:\vspace{-3pt}
$$
\PI(\zeta_0) = \max_{\zeta \, : \, d(\phi(\zeta),\phi(\zeta_0)) \leq S(\cell_0)}|\zeta-\zeta_0| .
$$
The point image is measured\cite{Harvey2011} as the product of the average receptive field size times the $\CMF$ at $\delta = 1$ degree\vspace{-3pt}
\begin{equation}\label{eq:pointimage}
\PI(\zeta_0) \approx \langle S\rangle_{\zeta_0} \, \CMF_1(\zeta_0) \ , \quad \langle S\rangle_{\zeta_0} = \frac{1}{H}\int_0^{H} S(\zeta_0,h) dh\vspace{-6pt} 
\end{equation}
which is consistent with the linear approximation (\ref{eq:CMF}).

\begin{figure}[h!]
\centering
\includegraphics[height=.14\textwidth]{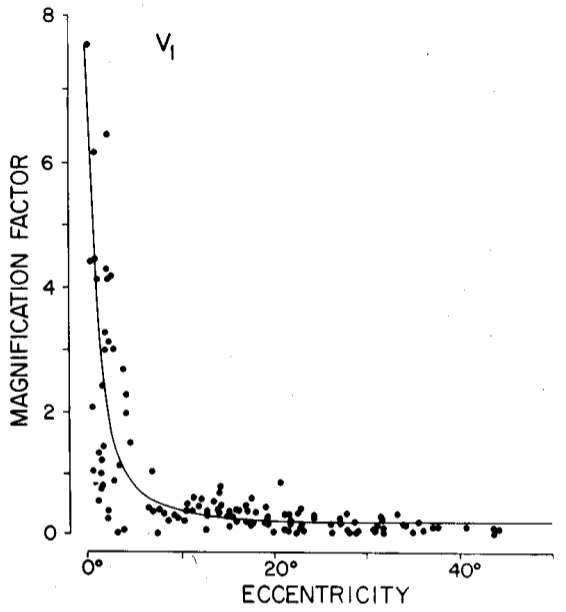} \hspace{16pt}
\includegraphics[height=.14\textwidth]{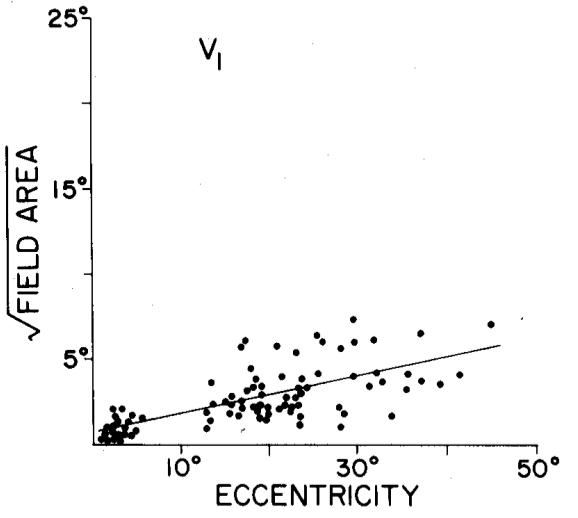} \hspace{40pt}
\includegraphics[height=.14\textwidth]{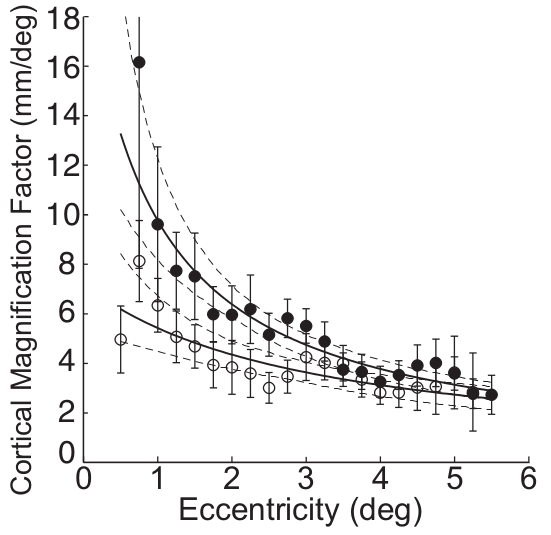} \hspace{16pt}
\includegraphics[height=.14\textwidth]{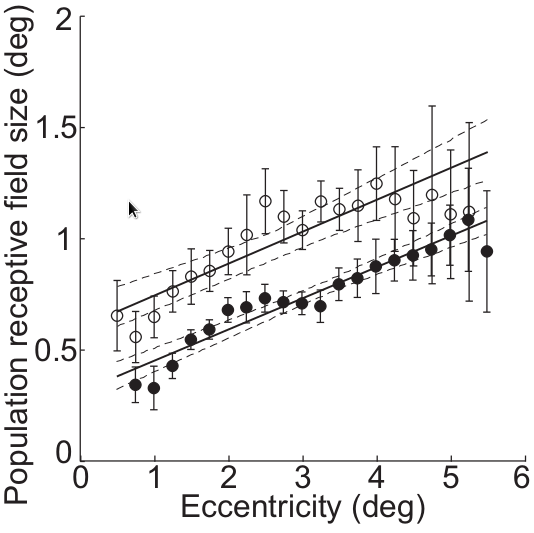}
\caption{Left: electrophysiological measurements in cats\cite{Gattass1981} of $\CMF$ and of the local average of $S$ against distance from the V1 point where the fovea is mapped (``eccentricity''). Right: fMRI measurements in humans\cite{Harvey2011}.}\label{fig:gattass}
\end{figure}

While $\phi$ and its derived quantity $\CMF$ depend only on the surface coordinate $\zeta$, it is not so for $S$, which actually depends both on $\zeta$ and on $h$: at any fixed $\zeta$ there are several scales $S$, which suggests that a \emph{multiscale analysis} is performed. On the other hand, if one considers the average $S$ over $h$ and looks at the variability with respect to the horizontal position $\zeta$, one sees that the size of receptive fields becomes larger as approaching the periphery of the visual field, at almost the same rate due to the log-polar map (see Figure \ref{fig:gattass}). This makes the point image an almost constant quantity over the whole V1 surface\cite{Harvey2011}. 

The log-polar mapping and the enlargement of the receptive fields towards the periphery of the visual field imply that the visual system performs a more refined analysis close to the fovea, losing precision towards the periphery. This is coherent also with the distribution of the density of retinal receptors, which is higher close to the fovea. The constancy of the point image implements the complementary principle of dedicating approximately the same amount of cortical tissue to the processing of local information over the whole visual field.

\newpage
\subsection{Orientations and shapes}

The majority of V1 neurons respond mostly to moving oriented stimuli, so their receptive profiles are more properly described by spatio-temporal filters\cite{DeAngelis1995, CBS}. Neurons with static behavior are called \emph{simple cells}, in contrast to spatio-temporal behaved neurons that belong to a class called \emph{complex cells}\cite{Carandini2005}. However, simple and complex cells have some common spatial characteristics \cite{CBS}, which consitute the only behaviors we will consider here.

Measurements of LNP receptive profiles\cite{Ringach2002} show very good fits with the so-called \emph{Gabor model}, that is given by a sinusoid oscillation under a Gaussian bell. They are defined on the retinal plane/visual field in terms of parameters $\beta \in [0,\pi/2)$, $\kappa \in \R^+$ and $\sigma = \binom{\sigma_1}{\sigma_2}$, where $\sigma_1, \sigma_2 \in \R^+$ as
$$
\begin{array}{rccl}
\g_{\beta,\kappa}^\sigma : & \R^2_\retina & \to & \R\\
& (x_1,x_2) & \mapsto & \displaystyle\frac{1}{\sqrt{\sigma_1 \sigma_2}}\cos(2 \pi \kappa x_1 + \beta) e^{-\frac{x_1^2}{2\sigma_1^2}} e^{-\frac{x_2^2}{2\sigma_2^2}}
\end{array}
$$
and are considered to be translated and rotated in all configurations on the visual field. More precisely, for $f : \R^2 \to \mathbb{C}$ let us denote with $T_q$ and $R_\theta$ translations and counterclockwise rotations on the plane
$$
\begin{array}{rcl}
R_\theta f(x) & = f(r_\theta x) \, , & \quad \theta \in S^1 = [0,2\pi) \ , \ \ \displaystyle r_{\theta} = \left(\begin{array}{cc}\cos\theta & -\sin\theta\\\sin\theta & \cos\theta\end{array}\right)\vspace{.5ex}\\
T_q f(x) & = f(x - q) \, , & \quad q \in \R^2_\retina .
\end{array}
$$
Any neuron located at $\cell \in V1$ is associated to a receptive profile $\psi_\cell(x) = T_q R_\theta \g_{\beta,\kappa}^\sigma (x)$, where the parameters $\beta$, $\kappa$, $\sigma$, $q$, $\theta$ depend on the neuron $\cell$. The dependence of $q$ on $\cell$ was discussed in the previous section. For $\sigma$, one can approximately consider the introduced quantity $S$ as corresponding to $2(\sigma_1+\sigma_2)$, since most of the power is contained within two standard deviations. The cortical map $\Theta$ accounts then for the dependence of $\theta$ on $\cell$.

With respect to $\beta$, a common simplification can be performed: since most V1 neurons\cite{Ringach2002} have a value of $\beta$ equal to $0$ or $\pi/2$, i.e. they are described either by a pure sine or a pure cosine\footnote{Traditionally, the cosine ones were called \emph{even cells} while the sine ones were called \emph{odd cells}.}, the cosine+phase term can be approximately substituted by a complex exponential without phase, meaning that both the real and the imaginary part provide a receptive profile. This produces the simplified model
\begin{equation}\label{eq:RP}
\begin{array}{rl}
\g_{\kappa}^\sigma = & \!\!\displaystyle\frac{e^{2 \pi \kappa x_1}}{\sqrt{\sigma_1 \sigma_2}} e^{-\frac{x_1^2}{2\sigma_1^2}} e^{-\frac{x_2^2}{2\sigma_2^2}}\vspace{1pt}\\
\psi_\cell(x) = & \!\!T_q R_\theta \g_{\kappa}^\sigma (x) = \displaystyle\frac{e^{2 \pi \kappa ((x_1-q_1)\cos\theta + (x_2-q_2)\sin\theta)}}{\sqrt{\sigma_1 \sigma_2}} e^{-\frac{((x_1-q_1)\cos\theta + (x_2-q_2)\sin\theta)^2}{2\sigma_1^2}} e^{-\frac{(-(x_1-q_1)\sin\theta + (x_2-q_2)\cos\theta)^2}{2\sigma_2^2}} .
\end{array}\vspace{-4pt}
\end{equation}

This model allows also to include easily a characteristic of complex cells. Their response is indeed generally well described by a slight modification of the LNP model, called the \emph{energy model}\cite{Carandini2005}, which consists of passing under the static nonlinearity not the linear filtering, but the square modulus of (\ref{eq:RP}), hence \emph{losing the phase}: an input stimulus $f : \R^2_\retina \to \R^+$ induces an activity on this type of neurons determined by a firing rate\vspace{-5pt}
$$
\mathscr{R}_\cell[f]=\sigmoid\Big(|\langle f, \psi_\cell\rangle_{\R^2_\retina}|^2\Big) .\vspace{-8pt}
$$

\begin{figure}[h!]
\centering
\includegraphics[height=.18\textwidth]{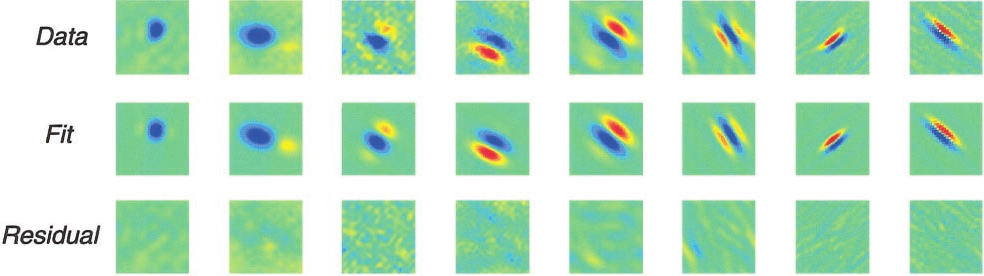}\hspace{40pt}
\caption{Data fit with Gaussian Gabors in macaque monkeys\cite{Ringach2002}.}\label{fig:Ringach2002_RFfit}
\end{figure}

\newpage

\subsection{On model parameters and topological dimensions}

The model of V1 receptive profiles given by (\ref{eq:RP}) is made of 6 parameters $\kappa, \theta, q_1, q_2$ and $\sigma_1, \sigma_2$, but V1 is only three dimensional, so only a subset of parameters is actually implemented. The bijection due to the retinotopic mapping allows to consider the parameter $q$ to depend approximately (up to the scatter) only on the two horizontal cortical coordinates $\zeta$. On the other hand, the evidence for orientation hypercolumns shown in Figure \ref{fig:bosking} allows us to consider the orientation parameter $\theta$, which depends essentially only on $\zeta$, to be a function of $q$. With respect to the spatial frequency parameter $\kappa$, experimental evidence for a columnar structure is less clear, as weakly ordered maps are observed from intrinsic signals\cite{Issa, Ribot} but their structure is debated\cite{Sirovich}. In any case the system of Gabor functions (\ref{eq:RP}) cannot make use of all frequencies at any point. Moreover, since the transversal dimension of V1 is finite, clearly also all scales can not be represented. If one wants to consider the linear transformation performed by V1 receptive profiles, which is the main contribution to the feedforward voltage of cells, as a generalized wavelet transform of an image, some special characterization of the space of parameters actually present in V1 is needed in order to understand its functional behavior and its computational purposes.

In order to reduce the problem, a useful simplification is that of disregarding the log-polar mapping and replace it with a linear mapping from the retina to the cortex. This can be considered as a local approximation, which amounts to forget about the center-periphery difference in resolution and to consider the $\CMF$ to be a constant. In order to keep a constant size for the cortical point image, the dependency of scales should consequently be reduced to the vertical coordinate $h$ only. From now on we will then consider the topographic mapping to be simply the identity, so that a first set of interparametric dependencies is provided by
\begin{equation}\label{eq:localfeatures}
q = \zeta \, , \ \theta = \Theta(q) \, , \ \sigma = \sigma(h) \, , \ \kappa = \kappa(q,h)
\end{equation}
where $q$ is then the independent coordinate on the cortical plane, which coincides with the image plane, while $h$ represents a transversal dimension defining the available scales over each position. The dependence of $\kappa$ is left for the moment free on $q$ and $h$. Moreover, by (\ref{eq:pointimage}), this implies that the point image can be approximately set to
\begin{equation}\label{eq:PI}
\PI \approx 2\langle \sigma_1 + \sigma_2\rangle = \frac{2}{H}\int_0^H (\sigma_1(h) + \sigma_2(h))dh
\end{equation}
i.e. the receptive field size is considered to be given approximately by 4 standard deviations.

An additional interparametric dependence is provided by the observation that spatial frequency and scale appear to be highly correlated. Several experiments have been performed in this respect, here we will focus on data of orientation selective receptive profiles from cats fitted with a Gaussian Gabor model\cite{CBS} and on previously cited data on macaques\cite{Ringach2002}, where the author has introduced the adimensional \emph{shape indices}\vspace{-4pt}
$$
n_i = \kappa\sigma_i \ , \ \ i = 1, 2 . \vspace{-4pt}
$$
The index $n_1$ measures the number of oscillations under the spatial window, while the ratio between the two indicates the anisotropy of the two dimensional Gaussian bell.

\begin{figure}[ht!]
\centering
\includegraphics[height=.24\textwidth]{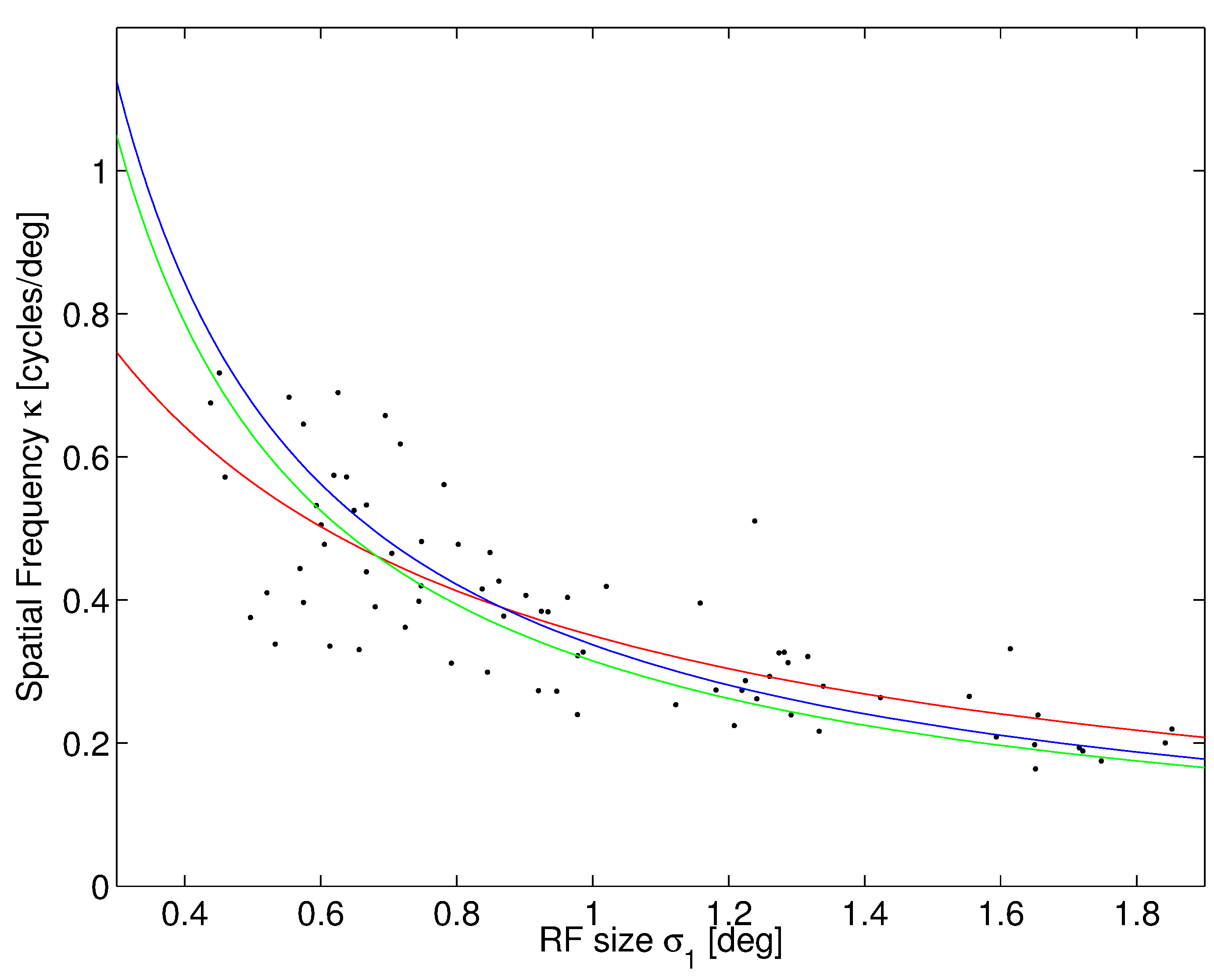}\hspace{20pt}
\includegraphics[height=.24\textwidth]{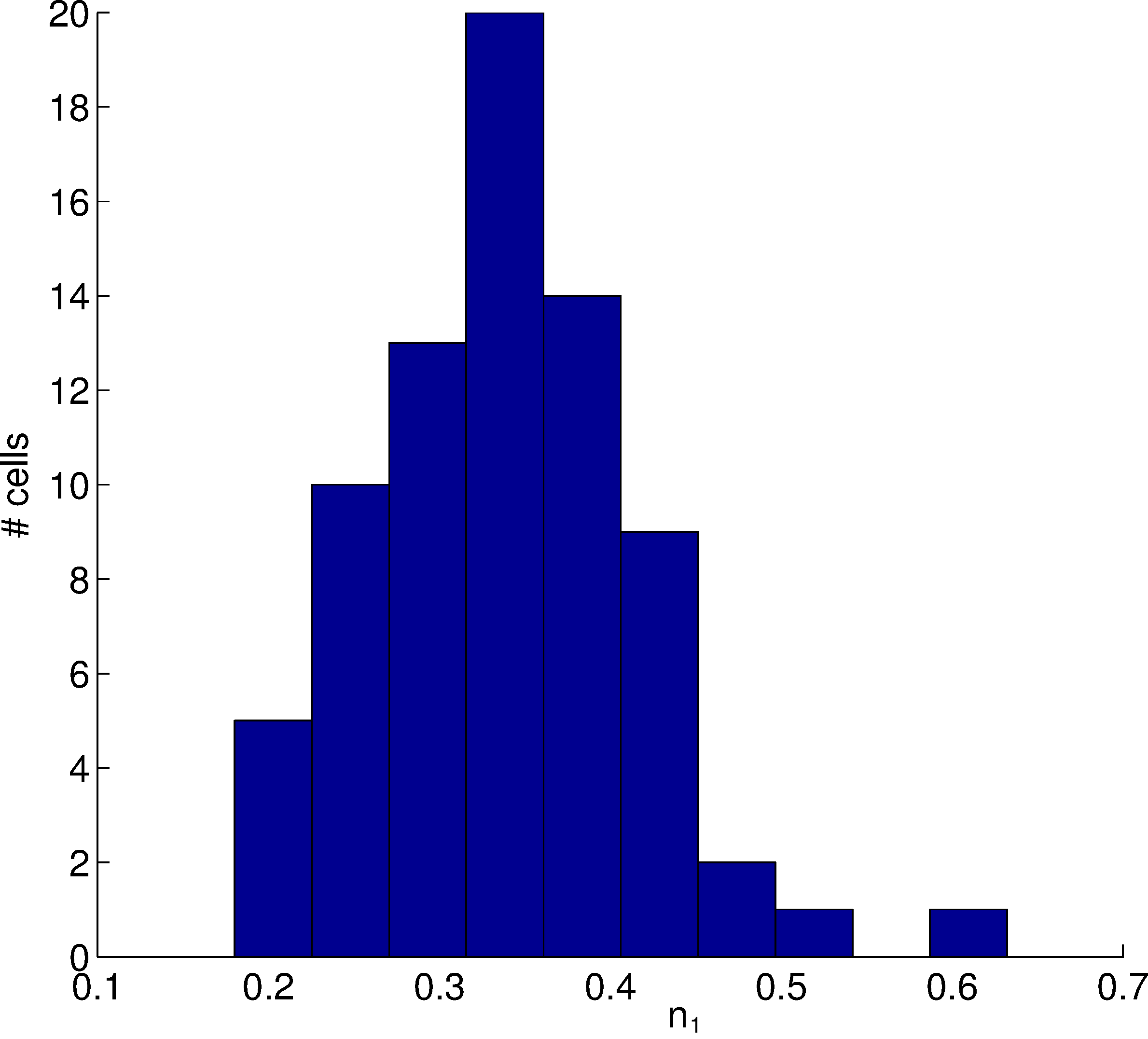}\hspace{40pt}
\includegraphics[height=.24\textwidth]{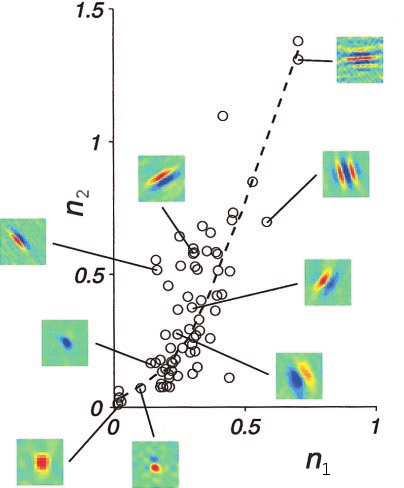}\hspace{20pt}
\caption{Left: fit with $\kappa = \frac{0.46}{\sigma_1 + 0.32}$ (red), with $\kappa = \frac{0.315}{\sigma_1}$ (green) and with $\kappa = \frac{\langle n_1\rangle}{\sigma_1}$ (blue) in cats\cite{CBS}. Center: histogram of $n_1$ (cats\cite{CBS}). Right: shape indices distribution in macaques\cite{Ringach2002}.}\label{fig:JOSAA}
\end{figure}

The relationship between spatial frequency $\kappa$ and scale $\sigma_1$ is shown in Figure \ref{fig:JOSAA}, left, as a scatter plot where each point represent a neuron. Among the simplest model functions, one of the best fits is provided by $\kappa = \frac{0.46}{\sigma_1 + 0.32}$ (red curve, $0.088$ root mean square error {\small{RMSE}}). A pure inverse proportionality provides a slightly poorer fit (green curve, 0.095 {\small{RMSE}}) with $\kappa \sigma_1 = 0.315$ which is about the mode of the distribution for the product $\kappa \sigma_1$; this differs slightly from the mean value $\kappa \sigma_1 = 0.34$ (blue curve, 0.99 {\small{RMSE}}). Nevertheless, due to the shape indications provided by $n_1$, it is relevant to observe in Figure \ref{fig:JOSAA}, center, how that shape index is distributed among the observed neurons. These are cells that have been chosen for being selective to orientation. This selection was not performed in the data displayed in Figure \ref{fig:JOSAA}, right, showing the relationship between the two shape indices. One can see the actual presence of a significative amount of cells with poor orientation selectivity, corresponding to low values of $n_1$. By computing the \emph{rotational uncertainty} of receptive profiles, it is indeed possible to classify cells whose value of $n_1$ is smaller than $0.16$ as unselective with respect to orientations\cite{BCS}. Moreover, one can see that the distribution of $n_1$ in macaques is very similar to that found in cats, as was already pointed out\cite{Ringach2002}.

We remark here that this correlation between receptive field size and spatial frequency, together with the variability of size with the transversal direction in the cortex, may be related to the apparently weak evidence of columnar organization of spatial frequencies\cite{Sirovich}.

Finally, it may be relevant to observe that the available data show evidence of a \emph{cutoff frequency}. Moreover, while the actual sizes and frequencies may vary among different species, the adimensional shape indices show an apparent universal behavior which is common at least to cats and macaques (and whose structure is not easily justified by means of standard optimization criteria \cite{Ringach2002}), with an approximate upper bound\vspace{-4pt}
\begin{equation}\label{eq:upperbound}
n_1 \leq \overline{n} \approx 0.5 .
\end{equation}

\section{Harmonic analysis and symmetries}

The linear behaviors of receptive field can be studied with instruments of harmonic analysis, focusing on the properties of the wavelet decomposition they perform on images and on the group symmetries associated to their shapes. This allows in particular to concretely formulate some open problems related to our visual system.

\subsection{Submanifolds in a group structure}

The set of receptive profiles can be fruitfully described, in an approximate way, in terms of the so-called \emph{affine Weyl-Heisenberg group}\cite{KT, AAG} in two dimensions. This is a semidirect product Lie group possessing a combined structure of time-frequency, scales and rotations. It reads $\G = \R^2_q \times \R^2_p \times \R_t \times S^1_\theta \times \R^+_a$, with composition law\vspace{-4pt}
$$
(q,p,t,\theta,a)\odot(q',p',t',\theta',a') = (q + ar_\theta^{-1}q', p + \frac{1}{a} r_\theta p', t + t' + a p\cdot r_\theta^{-1}q', \theta+\theta', aa') .\vspace{-4pt}
$$
Let us consider an approximate relation between the two scales as $\sigma_1 = \mu \sigma_2$ for a fixed value $\mu$, (Figure \ref{fig:JOSAA}, right, suggests a $\mu > 1$; the case $\mu = 1$ was treated in\cite{BCS}). Then, calling $\g_\mu(x) = \frac{1}{\sqrt{\mu}} e^{-\frac{x_1^2}{2}-\frac{x_2^2}{2\mu^2}}$, denoting with\vspace{-4pt}
\begin{equation}\label{eq:section1}
p = \kappa(\cos\theta,\sin\theta)\vspace{-2pt}
\end{equation}
and with $D_a f(x) = \frac{1}{a}f(\frac{x}{a})$ the ordinary two dimesional dilation operator, we have that (\ref{eq:RP}) reads
\begin{equation}\label{eq:RFrepresentation}
\psi_\cell(x) = T_{q} R_{\theta} M_{\binom{\kappa}{0}} D_{\sigma_1} g_\mu (x) = T_{q} M_p R_{\theta} D_{\sigma_1} g_\mu (x) = \Pi(q,p,\theta,\sigma_1)g_\mu (x).\vspace{-4pt}
\end{equation}
Now, $\Pi(q,p,\theta,a) = T_{q} M_p R_{\theta} D_a$ is a projective unitary representation of $\G$, because
\begin{align*}
\Pi(q,p,\theta,a)&\Pi(q',p',\theta',a') = T_{q} M_{p} R_{\theta} D_{a} T_{q'} M_{p'} R_{\theta'} D_{a'} = T_{q} M_{p} R_{\theta} T_{a q'} D_{a} M_{p'} R_{\theta'} D_{a'}\\
& = T_{q} M_{p} R_{\theta} T_{aq'} M_{\frac{1}{a} p'} D_{a} R_{\theta'} D_{a'} = T_{q} M_{p} T_{r_\theta^{-1}aq'} M_{r_\theta \frac{1}{a} p'} R_{\theta} R_{\theta'} D_{a} D_{a'}\\
& = e^{iap\cdot r_\theta^{-1}q'}T_{q} T_{r_\theta^{-1}aq'} M_{p} M_{r_\theta \frac{1}{a} p'} R_{\theta} R_{\theta'} D_{a} D_{a'} = e^{iap\cdot r_\theta^{-1}q'} \Pi(q + ar_\theta^{-1}q', p + \frac{1}{a} r_\theta p', \theta+\theta', aa') .
\end{align*}
The affine Weyl-Heisenberg group provides relevant symmetries for the analysis of multivariate data, as its structure contains as special cases the usual time-frequency, wavelet analysis and wave packets, as well as mixed behaviors \cite{KT, AAG}. It is well known that, in order to obtain resolutions of the identity, the whole group is too large and a proper subset needs to be considered. In the present case, a first reduction is provided by condition (\ref{eq:section1}), which amounts to align the direction of modulations with the rotations parameter. If the mother wavelet is isotropic (i.e. $\mu = 1$), this is equivalent to the elimination of the rotations, whose effect is totally incorporated in the changing of orientations of the modulation parameter (in (\ref{eq:RFrepresentation}) simply commute $R_\theta$ with $D_{\sigma_1}$ and use the rotational invariance of $g_1$). For anisotropic mother wavelets, relation (\ref{eq:section1}) constrains the dilation axes to be aligned with the reference frame provided by the modulation vector $p$.

\newpage

A second reduction in the parameters space $\G$ implemented by V1 can be ascribed to the approximate relation
\begin{equation}\label{eq:section2}
\kappa = K(\sigma_1) + \epsilon(q)
\end{equation}
showed in Figure \ref{fig:JOSAA}, left, where the function $K$ represents the fitting curve while $\epsilon$ accounts for the discrepancies, which give rise to the variability in the distribution of $n_1$. The models for $K$ given by the inverse proportionality $K(\sigma_1) = \frac{\alpha}{\sigma_1}$ or by the rational dependence $K(\sigma_1) = \frac{1}{\alpha\sigma_1 + \beta}$ are well-known relationships in the affine Weyl-Heisenberg group\cite{KT, AAG}. This amounts to model the linear filtering as a usual wavelet analysis (or a trivial modification of it) with a two-dimensional Morlet wavelet, whose role in image analysis for the detection of oriented features is well-known\cite{Lee, AMVA}. Since the data show that a relation $\kappa = K(\sigma_1)$ is not implemented exactly, this leaves space for some additional fine adjustments: the proposed dependence on $q$ represented by the $\epsilon$ term is suggested by the mentioned spatial frequency maps\cite{Issa, Sirovich, Ribot}.

Combining the constraints (\ref{eq:section1}) and (\ref{eq:section2}) with the ones given by (\ref{eq:localfeatures}), we can then approximately describe the parameter set $\X$ representing classical receptive fields of V1 as the subset of $\G$ given by
$$
\X = \bigg\{\cell = (q,p,\theta,\sigma_1) \in \G/\R_t \ \Big| \ 
\left\{
\begin{array}{rcl}
\theta & = & \Theta(q)\\
\sigma_1 & = & s(h)\\
p & = & \big(K(s(h)) + \epsilon(q)\big)\,(\cos\Theta(q),\sin\Theta(q))
\end{array}
\right. \ : \ q \in \R^2 \, , \ h \in [0, H]
\bigg\}
$$
where the function $\Theta:\R^2 \to S^1$ represents the orientation preference map and the function $s : [0,H] \to \R$ represents the variability of scales with the transversal cortical direction. With this approximation, the set of cortically implemented receptive profiles can then be written, for an appropriate parameter $\mu$, as $\{\Pi(\cell)g_\mu\}_{\cell \in \X}$.

\begin{figure}[h!]
\centering
\includegraphics[height=.175\textwidth]{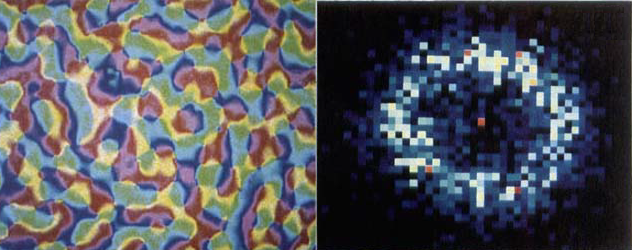} \hspace{20pt}
\includegraphics[height=.175\textwidth]{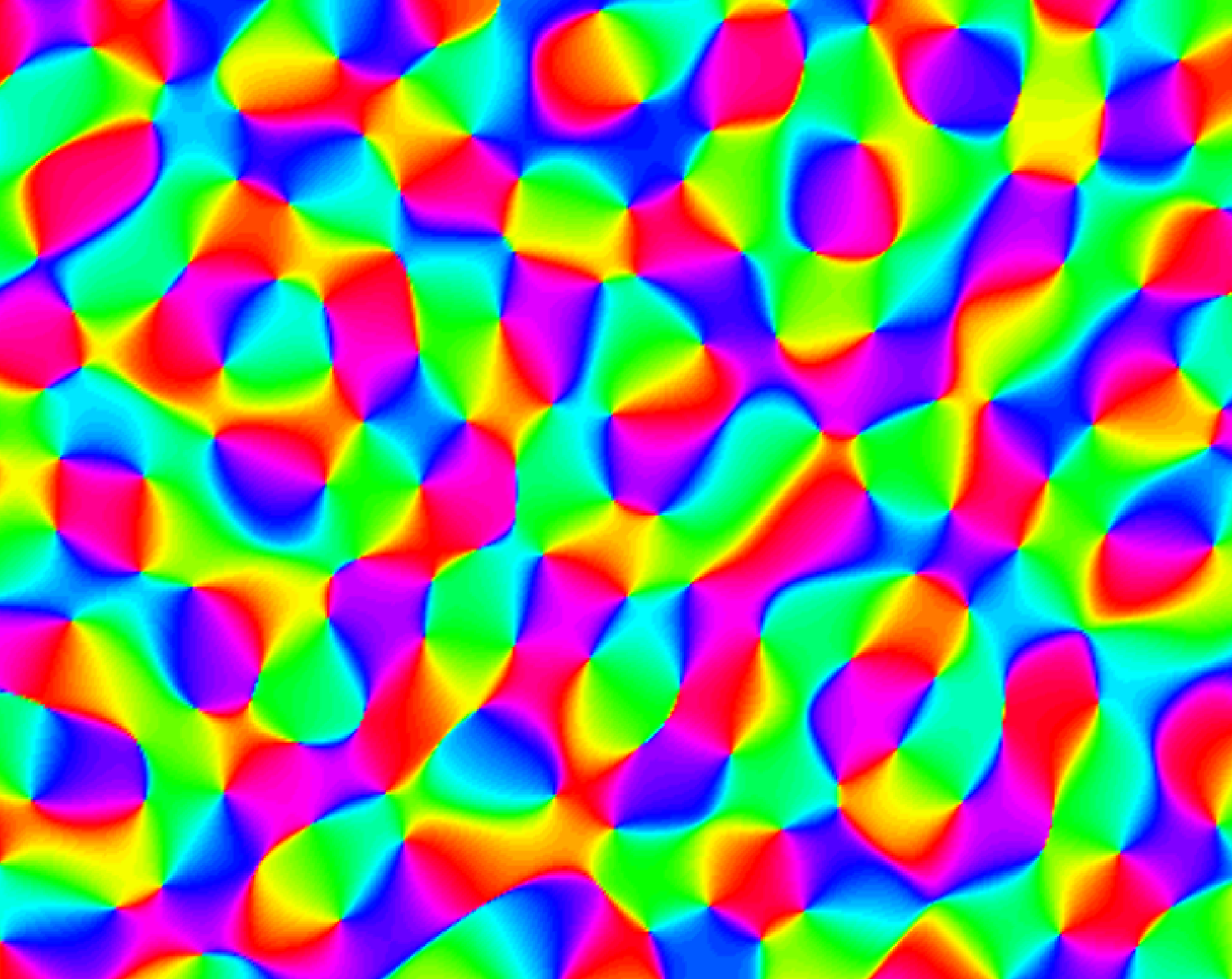}
\includegraphics[height=.12\textwidth]{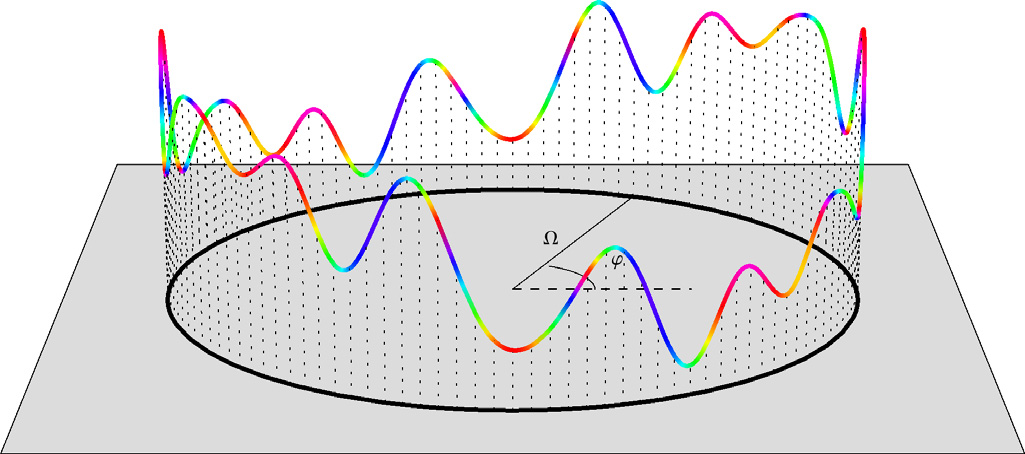}
\caption{Left: orientation preference map in macaque and its power spectrum\cite{models}. Right: map generated by equation (\ref{eq:pinwheels}) and an example of function $Z$ on the circle, where the height represents the modulus and the color represents the phase.}\label{fig:niebur}
\end{figure}

While there are apparently no models concerning the function $s$, which can probably be considered simply as a smooth bijection of $[0,H]$ onto some interval of scales $[\underline{s},\overline{s}]$, several models\cite{DurbinMitchinson, models, learnmap, BCSS, Kaschube} have been proposed for the map $\Theta$. A common one\cite{models} is based on the observation that the power spectrum of orientation preference maps is approximately concentrated on a thin annulus, where it shows and apparently random behavior. Thus, a simple way to obtain quasi-periodic orientation preference-like structures is given by
\begin{equation}\label{eq:pinwheels}
\Theta(q) = \frac{1}{2} \arg \int_0^{2\pi} e^{2\pi i \Omega (q_1 \cos\varphi + q_2 \sin \varphi)} Z(\varphi) d\varphi
\end{equation}
where $Z : [0,2\pi] \to \mathbb{C}$ is a highly irregular function (eventually a random process) and $\Omega$ is the radius of the circle in the Fourier plane, corresponding to the inverse of the typical correlation length, see Figure \ref{fig:niebur}. For what concerns spatial frequency maps and the term $\epsilon$, despite some evidence\cite{Hubener} of correlations with the map $\Theta$, they are still actively investigated\cite{Romagnoni}.

This class of maps is well studied in optics\cite{Berry}, in the present context they have been related to the irreducible representations of the group of rotations and translations\cite{BCSS}, which is the symmetry involved in the detection of orientations for the families (\ref{eq:RP}), and some of their geometric properties as random fields can be extended to other symmetries\cite{Afgoustidis}. By replacing the integral with a discrete sum (\ref{eq:pinwheels}), the associated quasi-periodic structure can also be obtained as results of nonlinear evolution equations\cite{Wolf} for the optimization of the so-called \emph{coverage-continuity} tradeoff, that is ``to preserve as far as possible neighborhood relations in parameter space''\cite{DurbinMitchinson}. Similar patterns were obtained by generative models via Hebbian learning\cite{learnmap}, and lateral connections seem to play a crucial role for their formation during animal development\cite{Kaschube}.

\newpage

A currently \underline{open problem}, which has been scarcely addressed in its full complexity\cite{Lee}, concerns the properties of the system $\{\Pi(\cell)g_\mu\}_{\cell \in \X}$ (or of even more realistic models of the implemented family of receptive profiles) considered as a wavelet analyzing family. In particular, it seems natural to ask what are its capabilities to represent images, or more in general what can be said about the function spaces associated to its linear decomposition. This appears to be a basic step to understand also the computational role of the several nonlinearities that intervene in the neural code with different modulations of this first linear processing\cite{Carandini2005}. It would also provide some clearer indications on the primal information collected by sensory area V1 and later processed by the different types of connectivities.

\subsection{Characteristic lengths and good coverage}

\begin{figure}[h!]
\centering
\includegraphics[height=.16\textwidth]{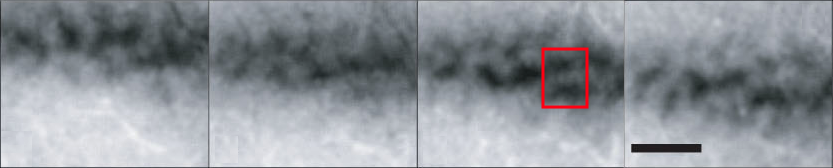}
\includegraphics[height=.16\textwidth]{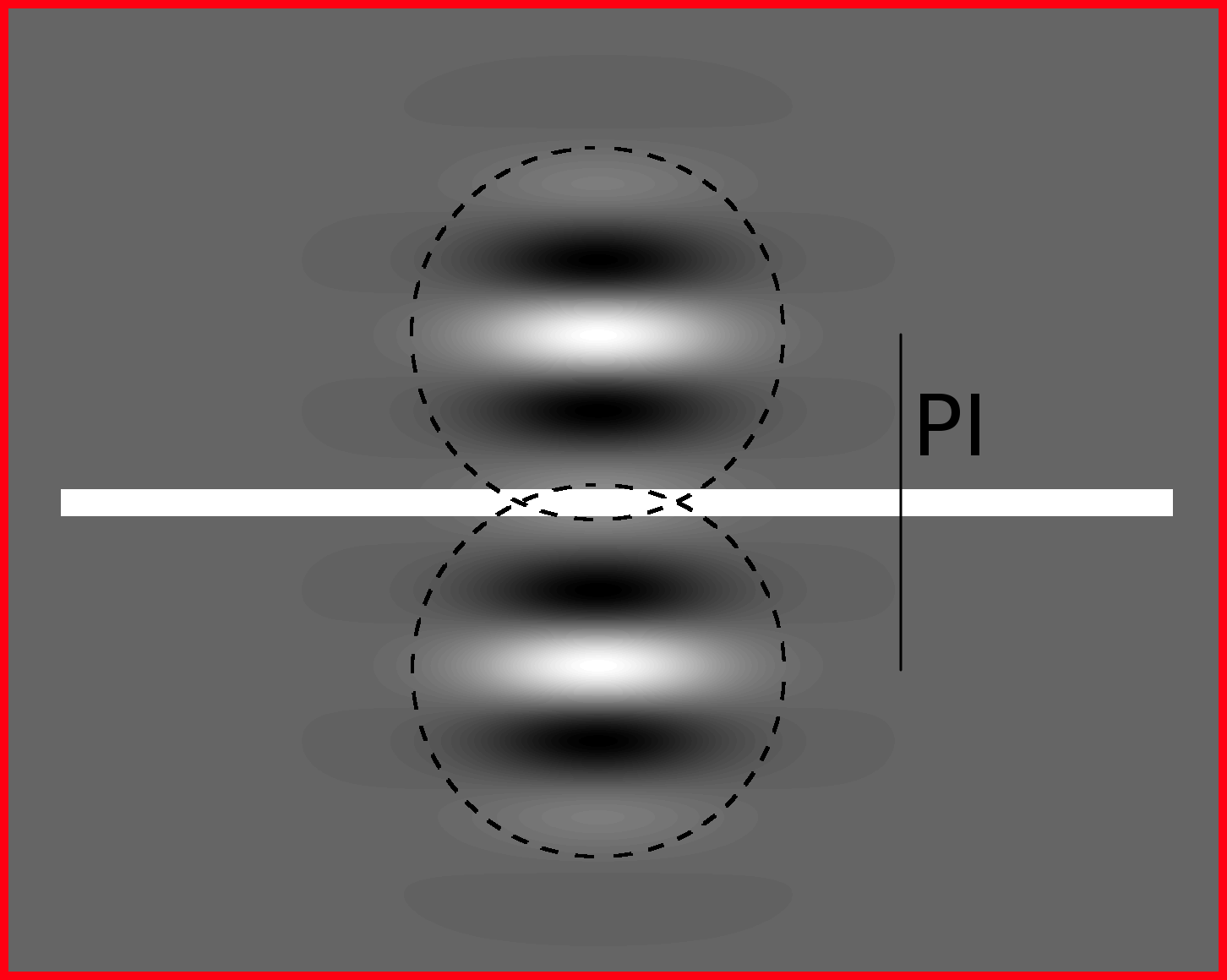}
\caption{Left: intrinsic signals of activity patterns in tree shrew's V1 produced by a thin line drifting in the visual space, at different times\cite{BCF2002} (scale bar, 1mm). Right: schematic picture of the two farthest neurons activated by the light bar.}\label{fig:Crowley}
\end{figure}

If one plans to perform a digital wavelet analysis in two dimensions with anisotropic wavelets, a natural approach may seem that of using all orientations over each point, i.e. to consider a stack of different oriented local measurements and to repeat it over the whole signal. However, visual cortex does not choose this strategy, since at each point $q$ of the image all the filters available (by changing the transversal cortical coordinate) have the same orientation, which changes as the (horizontal) position $q$ changes. A notion that tries to quantify the completeness of the representation of orientations is that of \emph{coverage}\cite{Swindale00}, and an optimization with respect to it means requiring that all orientations be represented over each point, by means of the overlapping of adjacent receptive fields. In a quasi-periodic configuration like the one of orientation preference maps, this condition implies that the length corresponding to the point image must be of the same order of the correlation length of these maps. Indeed, in that case all orientations would be represented in average within that distance, and all neurons lying within that distance would have receptive profiles that overlap at least partially. This necessary condition for coverage can then be summarized, in terms of the discussed models, by the condition
\begin{equation}\label{eq:constraint}
\PI \approx \frac{1}{\Omega}
\end{equation}
where $\Omega$ is the parameter in (\ref{eq:pinwheels}) and $\PI$ is given by (\ref{eq:PI}). This condition was actually checked experimentally by observing the activity produced by a thin line\cite{BCF2002}, see Figure \ref{fig:Crowley}. Such a stimulus can indeed activate neurons at a distance not larger than the point image, and only neurons selective to the same orientation will respond. This experiment shows then that the point image is approximately of the same size as the distance between two cortical regions whose neurons have the same preferred orientation. We observe also that condition (\ref{eq:constraint}) is actually a constraint that links the maps $s:[0,H] \to [\underline{s},\overline{s}]$ and $\Theta : \R^2 \to [0,\pi)$.

\subsection{Tuning curves and shape index cutoff}

Selectivity of V1 neurons to local frequencies is often measured in terms of \emph{orientation tuning} and of \emph{spatial frequency tuning} curves, which quantify the decay of neural response for stimulus parameters that differ from the preferred one\cite{Webster, Swindale98}. While this phenomenon can be partially explained in terms of classical receptive fields, it also offers a favorable experimental setting for checking for nonclassical effect, typically dynamical\cite{RingachTuning} and nonlinear/nonlocal\cite{Gardner, Sceniak} deformations. Here we simply observe that the qualitative behavior of these tuning curves is captured by the elements of the \emph{Gram matrix} of the system $\{\Pi(\xi)g_\mu\}_{\xi \in \X}$. Since this is meant to be only a rough approximation of the experimental results, for simplicity we will consider $\sigma_1 = \sigma_2 = a$ (i.e. $\mu = 1$), and denote with $\psi_{q,\kappa,\theta,a}(x) = \frac{1}{a}e^{2\pi i \kappa \big((x_1 - q_1)\cos\theta + (x_2 - q_2)\sin\theta\big)} e^{-\frac{|x - q|^2}{2a^2}}$, $q \in \R^2, \kappa \in \R^+, \theta \in S^1, a \in \R^+$, the receptive profiles.
\newpage

\begin{figure}[h!]
\centering
\includegraphics[height=.16\textwidth]{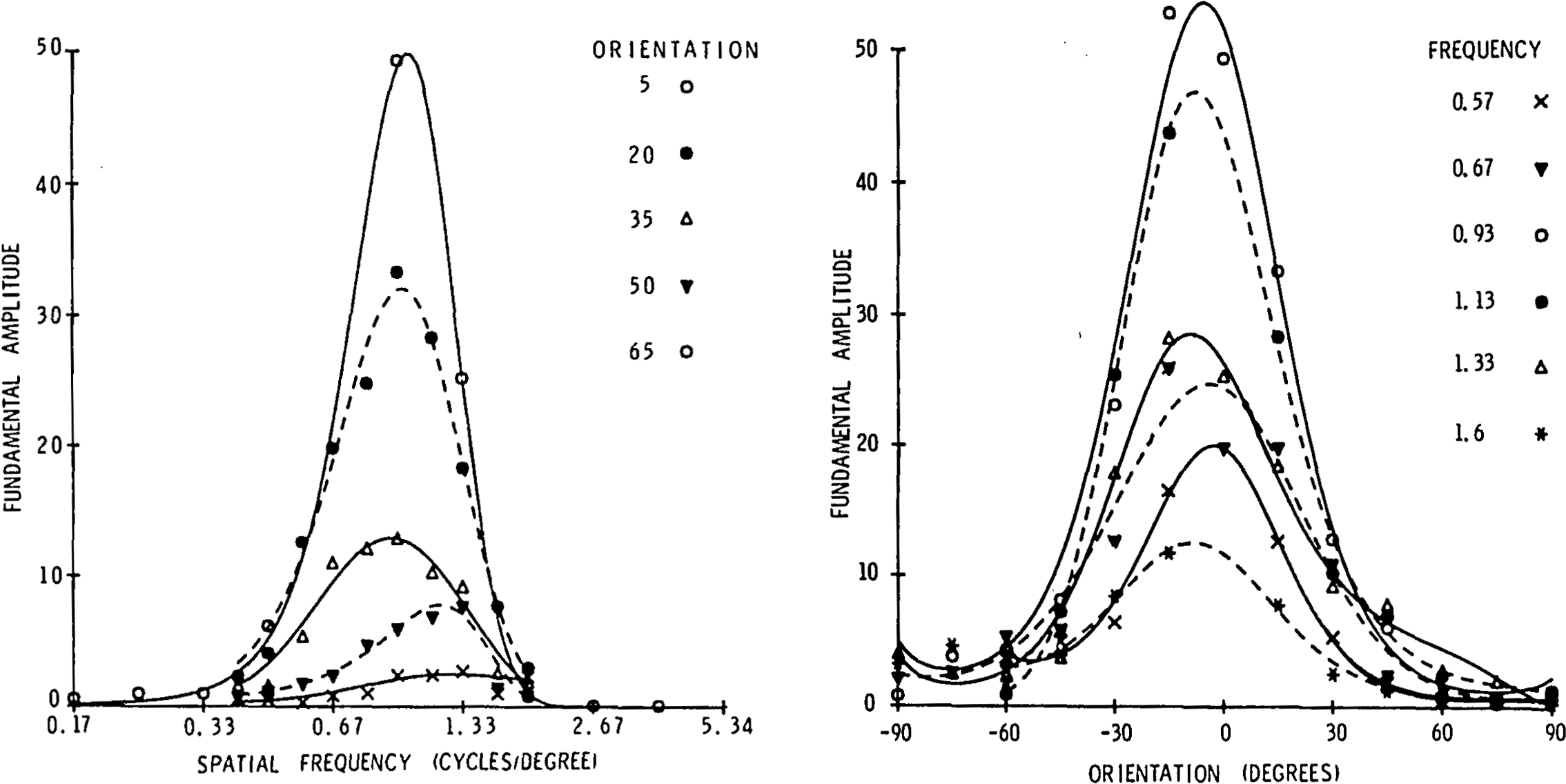} \hspace{20pt}
\includegraphics[height=.16\textwidth]{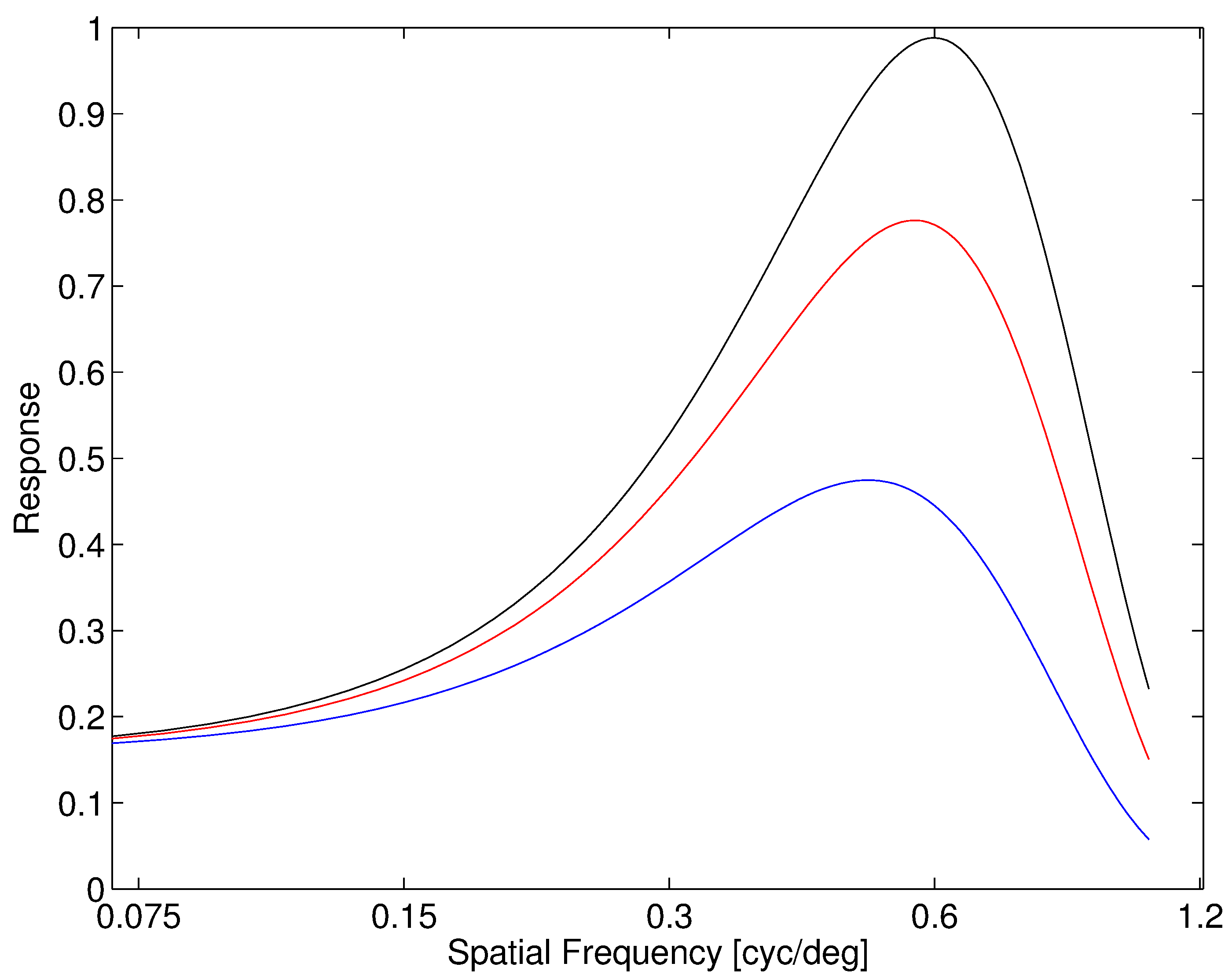}
\includegraphics[height=.16\textwidth]{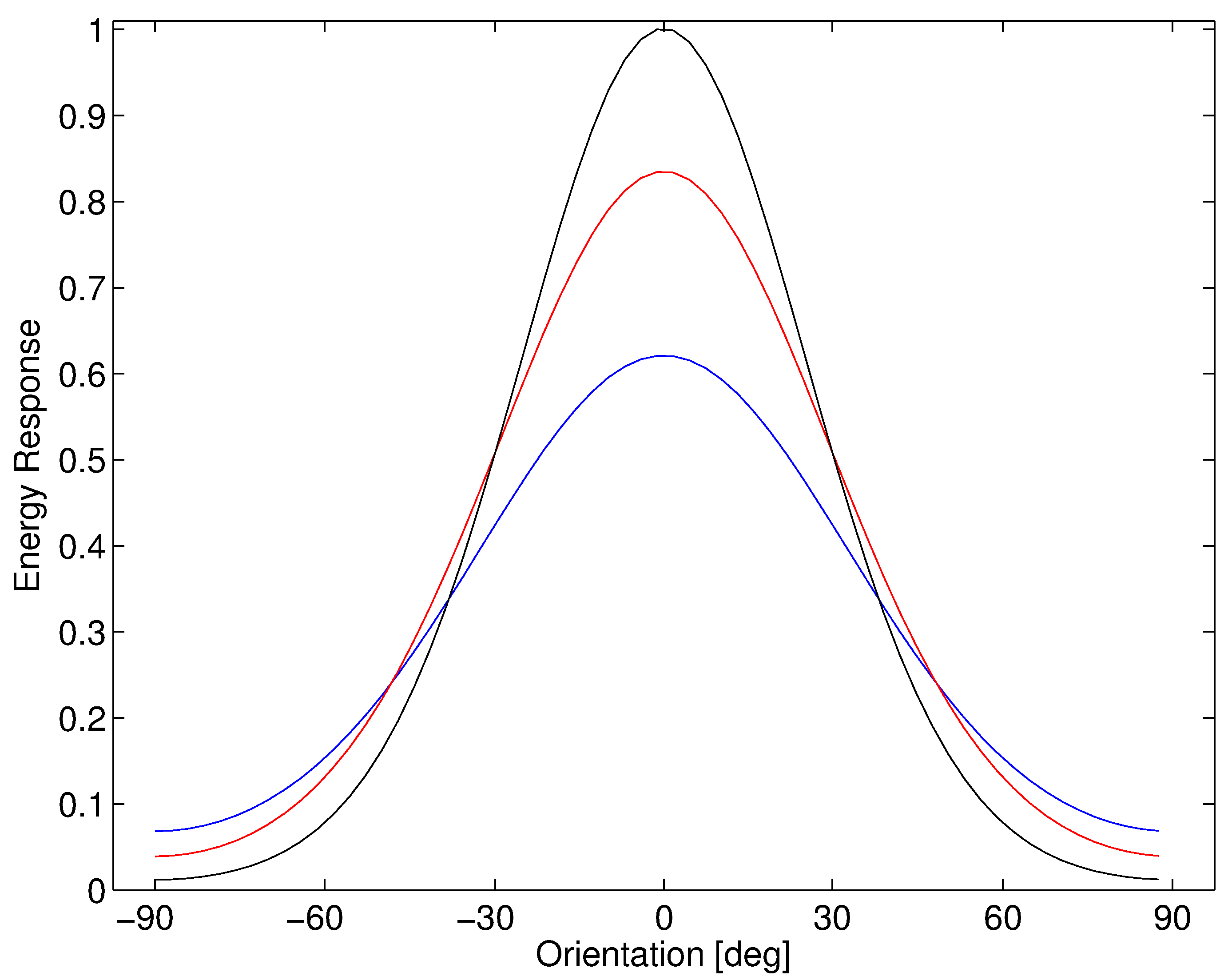}
\caption{Left: spatial frequency and orientation tuning curves for different values of the other parameter, in cats\cite{Webster}. Right: the function (\ref{eq:tuningcurve}) for a typical cell $\kappa_0 = 0.6, a = 0.6$, plotted with respect to $\kappa$ and $\theta$ for different values of the other parameter (normalized to have a maximum value 1). Observe that $\kappa$ is in logarithmic scale.}\label{fig:Webster}
\end{figure}

The square modulus of the Gram matrix for two elements at the same scale $a$ reads
\begin{equation}\label{eq:Gram}
\mathcal{G}(q,\kappa,\theta,a|q_0,\kappa_0,\theta_0,a) = \Big|\langle \psi_{q_0,\kappa_0,\theta_0,a},\psi_{q,\kappa,\theta,a}\rangle_{L^2(\R^2)}\Big|^2 = \pi^2 e^{-\frac{|q-q_0|^2}{2a^2}} e^{-2\pi^2a^2 \kappa_0^2} e^{-2\pi^2a^2(\kappa^2 - 2\kappa \kappa_0 \cos(\theta - \theta_0))}
\end{equation}
because (calling $p = \kappa(\cos\theta,\sin\theta)$)
$$
\bigg|\int_{\R^2} e^{-2\pi i p_0 x} e^{2\pi i p (x-q)}  e^{-\frac{|x|^2}{2a^2}} e^{-\frac{|x-q|^2}{2a^2}} \frac{dx}{a^2}\bigg| = \bigg|e^{-\frac{|q|^2}{4a^2}} \int_{\R^2} e^{-2\pi i (p_0 - p) x} e^{-\frac{|x-\frac{q}{2}|^2}{a^2}} \frac{dx}{a^2}\bigg| = \pi e^{-\frac{|q|^2}{4a^2}} e^{-\pi^2 a^2|p_0-p|^2} .
$$
The elements (\ref{eq:Gram}) can be considered as the energy of the response of a neuron to a stimulus shaped as a receptive profile. Since such a stimulus can provide the maximal linear response (diagonal elements) and is also parametrized by the features themselves, it offers a simple way to obtain tuning curves. The typical experimental setup for obtaining tuning curves consists of showing stimuli that produce an electrophysiologically measured response that does not depend on the absolute position of the neuron and is $\pi$-periodic on the angular variable. A simple way to obtain such a curve for a cell with preferred spatial frequency $\kappa_0$ and preferred orientation $\theta = 0$ is then via the following function
\begin{equation}\label{eq:tuningcurve}
t_{\kappa_0,a}(\kappa,\theta) = \mathcal{G}(0,\kappa,\theta,a|0,\kappa_0,0,a)^2 + \mathcal{G}(0,\kappa,\theta+\pi,a|0,\kappa_0,0,a)^2 = \pi^2 e^{-2\pi^2a^2 (\kappa_0^2+\kappa^2)} \cosh(4\pi^2a^2\kappa \kappa_0 \cos\theta).
\end{equation}
It qualitatively resembles well known models\cite{Swindale98} and shows the role of spatial frequency in the orientation tuning width. It is compared with actual measurements in cats\cite{Webster} in Figure \ref{fig:Webster}. The observed higher sharpness of the real tuning curves is thought to be related to nonlinear/nonlocal additional cortical mechanisms\cite{Gardner, Sceniak}.

The elements (\ref{eq:Gram}) can also be used to obtain a \emph{relationship between the shape index cutoff and the characteristic length} (\ref{eq:constraint}), by an argument introduced in\cite{BCS} concerning correlations. The function
\begin{equation}\label{eq:correlations}
\C_{\kappa,a}(q,\theta) = \mathcal{G}(q,\kappa,\theta,a|0,\kappa,0,a)^2 + \mathcal{G}(q,\kappa,\theta+\pi,a|0,\kappa,0,a)^2 = \pi^2 e^{-4\pi^2 n^2} e^{-\frac{|q|^2}{2a^2}} \cosh(4\pi^2 n^2\cos\theta)
\end{equation}
\begin{wrapfigure}{r}{.18\textwidth}
\centering
\includegraphics[height=.18\textwidth]{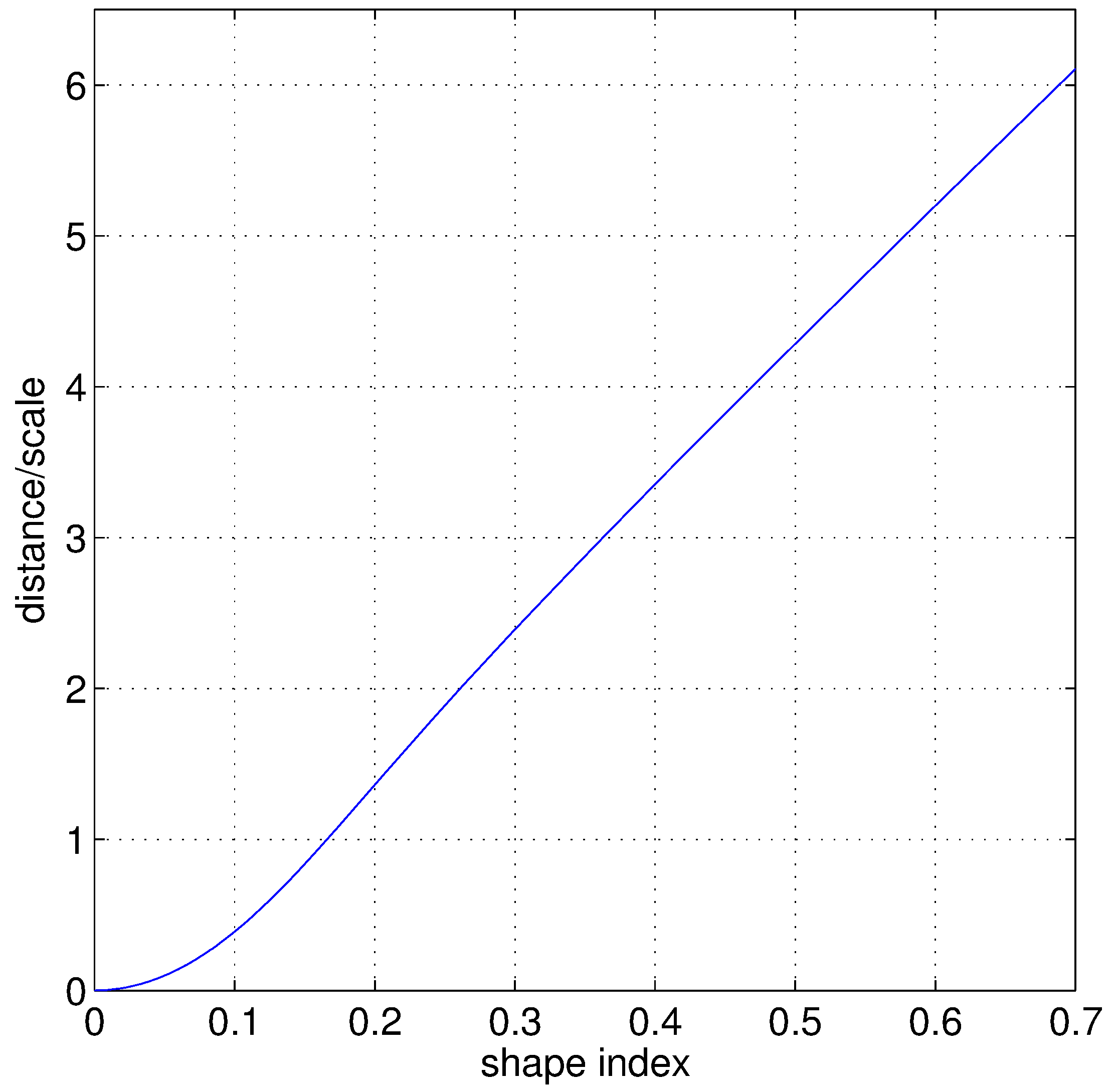}
\caption{$\lambda$ vs $n$ plot.}\label{fig:shape}
\end{wrapfigure}
can indeed be considered as a measurement of \emph{energy autocorrelation} in space and orientation for a receptive profile having spatial frequency $\kappa$ and scale $a$, denoting with $n = \kappa a$ its (isotropic) shape index. Since an orientation difference of $\theta = \frac{\pi}{2}$ provides the minimum correlation that can be attained due to the orientation selectivity mechanism only, it may be relevant to observe what is the spatial distance at which a receptive profile has the same correlation due to spatial decay only. By (\ref{eq:correlations}) one obtains
$$
\C_{\kappa,a}(q,0) = \C_{\kappa,a}(0,\frac{\pi}{2}) \ \Leftrightarrow \ \lambda = \frac{|q|}{a} = \sqrt{2\log(\cosh(4\pi^2 n^2))} .
$$
The dependency of the ratio $\lambda = \frac{|q|}{a}$ on the shape index $n$ is shown in Figure \ref{fig:shape}. It is possible to observe that, for a value of $n$ around $0.5$, which coincides approximately with the cutoff bound (\ref{eq:upperbound}) one obtains a distance that is approximately 4 times the scale, providing an effective size of a receptive profile corresponding to what is inside 2 standard deviations. By (\ref{eq:PI}), this was considered to correspond approximately to the point image, and by (\ref{eq:constraint}) this length is apparently mapped on V1 as the characteristic length of the quasi-periodic map $\Theta$.

This can be interpreted as a principle of \emph{dimensionality reduction constrained to optimally independent representation of orientations}, in the following sense. As previously observed, V1 does not have a sufficiently high topological dimension to implement all scales/frequencies and orientations over each point, so it has adopted compromises such as the one described by the coverage strategy. On the other hand, if we measure the independence of receptive profiles in terms of correlations, then the maximal independence with respect to orientations can be obtained equivalently by translations at a given distance. This distance can be estimated to grow with the shape index, as described by Figure \ref{fig:shape}, and reaches the characteristic length of the orientation preference maps at about the cutoff value for the shape index. At that distance, two receptive profiles can then be considered as collecting a sufficiently independent information that justifies a repetition of a new full set of orientations. Say it from another point of view, orientation preference maps may be a way to map a compact variable on the highly redundant sampling space of positions. However, spatial frequency variable is not compact, so in order to be projected down one needs a cutoff. The one provided by (\ref{eq:upperbound}) is consistent with the purpose of having maximal decorrelation at the distance corresponding to the quasi-periodicity of the map $\Theta$. Finally, we observe that frequency cutoff could be related to the finite resolution at the retinal level\cite{marr}, but due to the intermediate optic nerve information compression and to the LGN preprocessing\cite{Valois} this dependence may be highly nontrivial.

\section{Conclusions}

Even if much is known about the computations performed by primary visual cortex V1, still many problems concerning the fine structure of these computations as well as their purposes in the processing of visual information are open. This paper has focused on a restricted class of behaviors concerning classical receptive fields, which provide a primal linear response of V1 neurons to visual stimuli that is later processed by nonlinear mechanisms and by dynamical lateral and feedback connectivities. Many studies approach the computation implemented by V1 as being optimized for best information detection (even if other purposes have been proposed, such as that of maximizing inference\cite{Friston} via Bayesian\cite{Tonda} schemes), and we have provided evidence that some of the design principles encountered in V1 are actually compatible with this view. The presented approach focuses on the geometry of classical receptive fields, by considering their contribution to neural code as a linear wavelet analysis of visual stimuli. Due to the prominent role of selectivity for local orientation, the substructure given by the group of translations and rotations has been often considered in relation with with the cortical two dimensional space-frequency analysis\cite{BC, SCP}, as well as with the modeling of contour perception and of lateral connectivities\cite{CS, SCS, gauge}, and it has provided a fruitful environment for cortical-inspired image processing on groups\cite{Duits2013, Duits2015}. More in general, the geometry associated to V1 receptive profiles has provided a well-established set of models of the cortical architecture\cite{CSbook}, and is thought to play a key role in the constitution of perceptual units\cite{SCpercepts, CBCS}. Most of this geometry can be obtained by experimental setups that make use of parametric sets of simple stimuli, but the more recent approach to the study of receptive fields is actually based on the exploitation of nonlinearities and on the use of natural stimuli\cite{Felsen}. This provides indeed the framework for obtaining families of receptive field-like linear analyzers in terms of optimality criteria extracted from the statistics of natural images \cite{Olshausen, VinjeGallant}, and to reconcile some behaviors observed in V1 with recent approaches to high dimensional statistics, such as compressed sensing and sparsity\cite{Fletcher, Ganguli}. On the other hand, still much of the linear behavior continues to provide unsolved problems form the point of view of the strategies used by V1 for collecting informations. We have mentioned that of characterizing the space of signals that can be represented by V1 classical receptive fields, but also the appearance of \emph{quasi-periodic structures} in the space of coefficients of a Gabor-like system seems to be suggestive in view of recent advances on similar approaches to sampling problems\cite{matei}. Moreover, the energy model for complex cells activity should be actually related to issues of \emph{phase retrieval}\cite{phase}. Finally, the way nonlinearities intervene into the neural coding seems to be quite unusual with respect to the mainstream approaches to nonlinear approximation, as some of them seem to act mainly as \emph{deformations of the linear behavior}, with frequent evidence of adaptivity mechanisms. Few theoretical studies in approximation theory deal with similar kinds of nonlinear behaviors\cite{Sigl}, which however could provide effective alternative strategies from the ones presently considered in signal processing.

\end{document}